\documentclass[aps,prl,twocolumn,superscriptaddress,10pt,article,nofootinbib,showpacs,longbibliography]{revtex4-1}
\usepackage{amsfonts}
\usepackage{amsmath}
\usepackage{amsthm}
\usepackage{braket}
\usepackage{graphicx}
\usepackage{hyperref}
\hypersetup{
    colorlinks=true,
    linkcolor=blue,
    filecolor=cyan,      
    urlcolor=magenta,
    citecolor=purple,
}
\usepackage[dvipsnames]{xcolor}
\usepackage{amssymb}
\usepackage{dsfont}
\usepackage{verbatim}
\usepackage{amsmath,amscd}
\usepackage{multirow}
\usepackage[normalem]{ulem}
\usepackage{bm}
\def \k{{\mathbf k}}
\def \q{{\mathbf q}}

\def \A{{\mathbf A}}

\def \n{{\mathbf n}}
\def \j{{\mathbf j}}
\def \r{{\mathbf r}}

\def \A{{\mathbf A}}
\def \B{{\mathbf B}}
\def \E{{\mathbf E}}

\def \beq{\begin{eqnarray}}
\def \eeq{\end{eqnarray}}

\DeclareMathOperator{\Tr}{Tr}

\begin{document}

\title{Single-spin qubit magnetic spectroscopy of two-dimensional superconductivity}

% \author{Authors}
\author{Shubhayu~Chatterjee}
\thanks{S.C. and P.E.D. contributed equally to this work.}
\affiliation{Department of Physics, University of California, Berkeley, CA 94720, USA}
\author{Pavel~E.~Dolgirev}
\thanks{S.C. and P.E.D. contributed equally to this work.}
\affiliation{Department of Physics, Harvard University, Cambridge, Massachusetts 02138, USA}
\author{Ilya~Esterlis}
\affiliation{Department of Physics, Harvard University, Cambridge, Massachusetts 02138, USA}
\author{Alexander~A.~Zibrov}
\affiliation{Department of Physics, Harvard University, Cambridge, Massachusetts 02138, USA}
\author{Mikhail~D.~Lukin}
\affiliation{Department of Physics, Harvard University, Cambridge, Massachusetts 02138, USA}
\author{Norman~Y.~Yao}
\affiliation{Department of Physics, University of California, Berkeley, CA 94720, USA}
\affiliation{Materials Sciences Division, Lawrence Berkeley National Laboratory, Berkeley CA 94720, USA}
\author{Eugene~Demler}
\affiliation{Department of Physics, Harvard University, Cambridge, Massachusetts 02138, USA}
\affiliation{Institute for Theoretical Physics, ETH Zurich, 8093 Zurich, Switzerland}

\begin{abstract}
A single-spin qubit placed near the surface of a conductor acquires an additional contribution to its $1/T_1$ relaxation rate due to magnetic noise created by electric current fluctuations in the material. We analyze this technique as a wireless probe of superconductivity in atomically thin two-dimensional materials. At temperatures $T \lesssim T_c$, the dominant contribution to the qubit relaxation rate is due to transverse electric current fluctuations arising from quasiparticle excitations. We demonstrate that this method enables detection of metal-to-superconductor transitions, as well as investigation of the symmetry of the superconducting gap function, through the noise scaling with temperature. 
We show that scaling of the noise with sample-probe distance provides a window into the non-local quasi-static conductivity of superconductors, both clean and disordered. 
At low temperatures the quasiparticle fluctuations get suppressed, yet the noise can be substantial due to resonant contributions from collective longitudinal modes, such as plasmons
in monolayers and Josephson plasmons in bilayers.
Potential experimental implications are discussed.
\end{abstract}

\maketitle

\emph{Introduction---} A superconductor is a phase of matter characterized by the dissipation-free flow of electrical current owing to the intrinsic quantum coherence between electron pairs~\cite{Tinkham}. The recent discovery of robust superconductivity in a variety of two-dimensional (2D) materials, such as moir\'{e} graphene \cite{cao2018,lu2019,yankowitz2019,park2021,hao2021}, transition metal dichalcogenides (TMDs) \cite{shi2015} and monolayer FeSe \cite{ge2015FeSe,Wang2012FeSe,Song2011FeSe}, has spawned immense theoretical and experimental interest. 
The experimental tunability of 2D materials, e.g., in situ varying of carrier density via gate voltages, bandwidth using displacement fields, and dielectric properties/screening with substrates, makes them highly relevant for practical applications. 
However, while superconductivity in bulk solids and thin films has been extensively studied \cite{Tinkham}, characterizing it in bona fide 2D materials poses a significant challenge. 
In certain materials like TMDs, superconductivity itself can be hard to detect due to the difficulty of making contacts for transport experiments \cite{Allain2015}.
Conventional bulk probes of the nature of the superconducting gap, such as specific heat or thermal conductivity, are dominated by the contribution from the substrate. 
Local probes like STM, being particularly sensitive to inhomogeneities in the sample, are often inconclusive about the symmetry of the gap function \cite{Gu2020}.
Measurements of the Meissner effect are also challenging in atomically thin superconductors and require local magnetometry~\cite{zhang2015onset,xu2019mapping}. 
This calls for new experimental probes which can be used to diagnose the onset of superconductivity and elucidate the pairing symmetries in 2D materials. 

\begin{figure}[!htbp]
    \centering
    \includegraphics[width=0.48\textwidth]{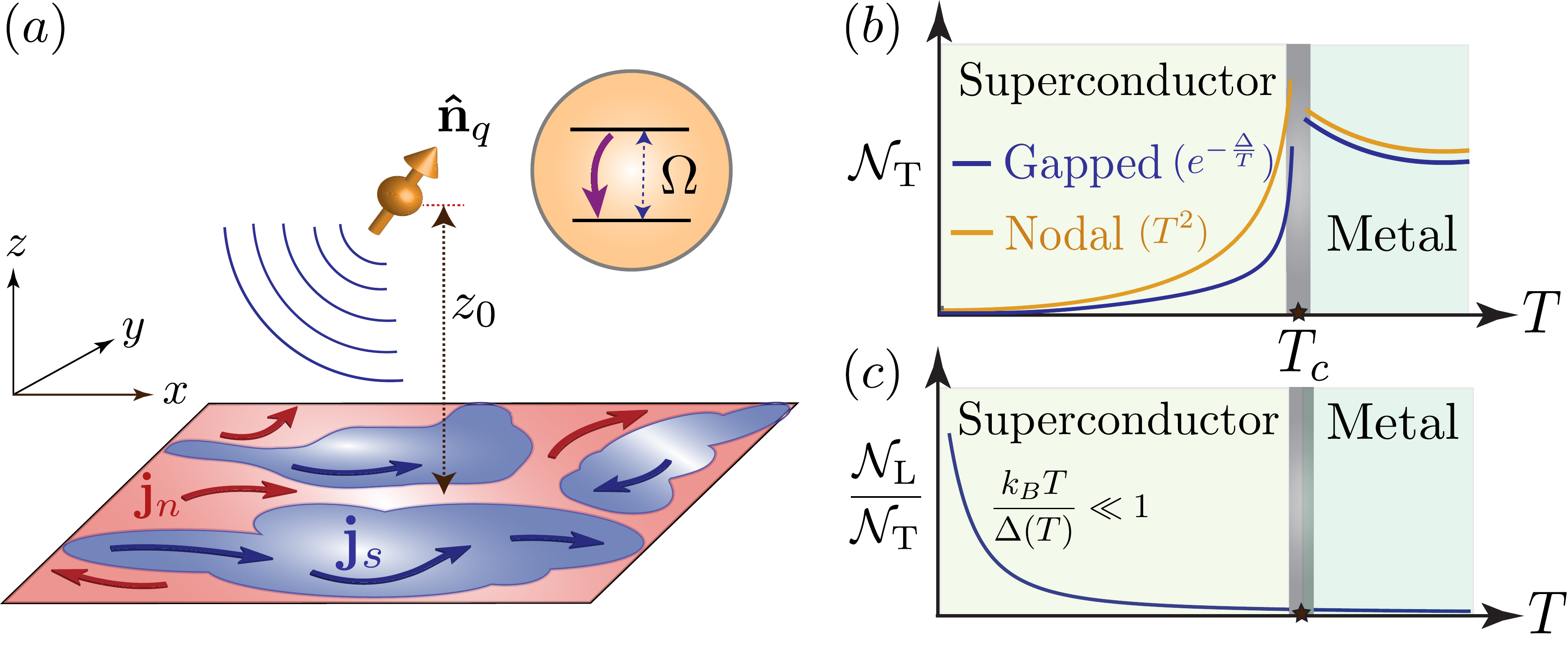}
    \caption{(a) Schematic of the set-up, with qubit at a distance $z_0$ from a 2D superconducting sample. The qubit polarization decays due to magnetic field noise created by fluctuating normal currents $\j_n$ and super-currents $\j_s$ in the film. (b) Drop in the transverse noise $\mathcal{N}_{\rm T}$ when $T$ falls below $T_c$ for gapped and nodal superconductors (excluding effects of critical dynamics in narrow gray window around $T = T_c$). (c) Schematic of $\mathcal{N}_{\rm L}/\mathcal{N}_{\rm T}$ as a function of $T$, showing large enhancement at low $T$, making it possible to detect longitudinal collective modes.}
    \label{fig:SchematicSummary}
\end{figure}

In this Letter, we propose quantum noise spectroscopy by impurity spin qubits, such as nitrogen-vacancy (NV) centers in diamond, as a natural non-invasive wireless probe of 2D superconductivity. 
The probe qubit is initialized in a fully polarized state at a distance $z_0$ above our 2D sample [Fig.~\ref{fig:SchematicSummary}(a)]. 
Coupling to the noisy magnetic field created by thermal current fluctuations in the sample causes the qubit polarization to decay. 
The decay-rate ($1/T_1$), studied as a function of experimentally tunable parameters such as qubit-probe distance $z_0$, probe frequency $\Omega$, and temperature $T$, furnishes valuable information about the nature of superconductivity in the 2D sample.  

Specifically, we predict that the sharp reduction of noise due to suppression of transverse current fluctuations, stemming from the onset of a superconducting gap, allows one to detect the phase transition from a normal metal to a superconductor [Fig.~\ref{fig:SchematicSummary}(b)]. 
Second, we demonstrate that the nature of the superconducting gap (nodal vs non-nodal) can be deciphered by studying the noise as a function of temperature. 
Simultaneously, the scaling of the noise with probe-qubit distance can be used to study the non-local conductivity in the quasi-static limit ($\q \neq 0, \Omega \to 0$), a regime complimentary to existing probes (such as dc transport, THz spectroscopy, etc).
We elucidate these distinct scaling regimes in both clean and disordered superconductors, carefully accounting for additional modifications arising from super-flow. 
Third, we illustrate that deep in the superconducting phase the noise is dominated by longitudinal current fluctuations [Fig.~\ref{fig:SchematicSummary}(c)], in sharp contrast to metals \cite{Agarwal2017}. 
This owes its origin to the suppression of thermally excited quasiparticles at low temperatures, and allows us to detect collective longitudinal modes, such as gapless plasmons in monolayers and gapped Josephson plasmons in weakly interacting bilayer superconductors \cite{PhysRev.112.1900,kulik1974josephson,barone1982physics,PhysRevB.50.12831,Sun}. 
We conclude by providing realistic noise estimates for materials of recent interest, and argue that such measurements lie within experimental reach.
%the reach of current experiments. 

\emph{Set-up and model---} To detect magnetic noise, we consider %a simple geometry: 
an isolated impurity qubit placed at $\r_0 = (0,0,z_0)$, i.e, 
at a distance $z_0$ above a homogeneous 2D sample in the x-y plane (Fig.~\ref{fig:SchematicSummary}a). 
The qubit Hamiltonian includes an intrinsic level-splitting $\Omega$ along quantization axis $\hat{\n}_q$, and a linear coupling to the local magnetic field $\B(\r_0,t)$ (henceforth $\hbar=1$):
\beq
H_q =  \frac{\Omega}{2}(\hat{\n}_q \cdot \bm{\sigma}) + g \mu_B \B(\r_0, t) \cdot \bm{\sigma}.
\eeq
The depolarization rate of the qubit is directly proportional to the magnetic noise at the qubit location. 
This noise arises predominantly due to current fluctuations in the sample, assumed to be at thermal equilibrium at temperature $T$ \cite{Langsjoen}. By analyzing the relaxation rate of the qubit for different orientations of the quantization-axis, one can isolate different components of the magnetic noise tensor $\mathcal{N}_{ab}(\Omega) \equiv \frac{1}{2} \int_{-\infty}^\infty dt \, e^{i \Omega t} \langle \{ B_a(\r_0,t), B_b(\r_0, 0) \} \rangle_T$ \cite{Agarwal2017}.
These in turn can be used to investigate two qualitatively different types of noise:
(i) transverse noise $\mathcal{N}_{\rm T}$, arising from transverse current fluctuations ($\j_{\rm T}(\q) \perp \q$) that do not create charge imbalance in the sample, is related to the reflection coefficient  $r_s(\q,\Omega)$ for s-polarized electromagnetic (EM) waves; (ii) longitudinal noise $\mathcal{N}_{\rm L}$, arising from longitudinal current fluctuations ($\j_{\rm L}(\q) \parallel \q$), is related to the reflection coefficient $r_p(\q,\Omega)$ for p-polarized EM waves ($\q$ is the in-plane momentum).
Reference~\cite{Agarwal2017} argued that the latter can be ignored in metals because p-polarized waves create charge density modulations that are disfavored by long-range Coulomb interactions. However, the presence of superconductivity can suppress $r_s$ at low temperatures and thereby provide a gateway to probe $r_p$.
In what follows, we characterize magnetic noise due to transverse and longitudinal current fluctuations in 2D superconductors and determine physical regimes where each contribution dominates. Here, we present our main physical results, but relegate computational details and additional discussion to a complementary paper~\cite{Shubhayu2}. 

\begin{figure}
    \centering
    \includegraphics[width=0.4\textwidth]{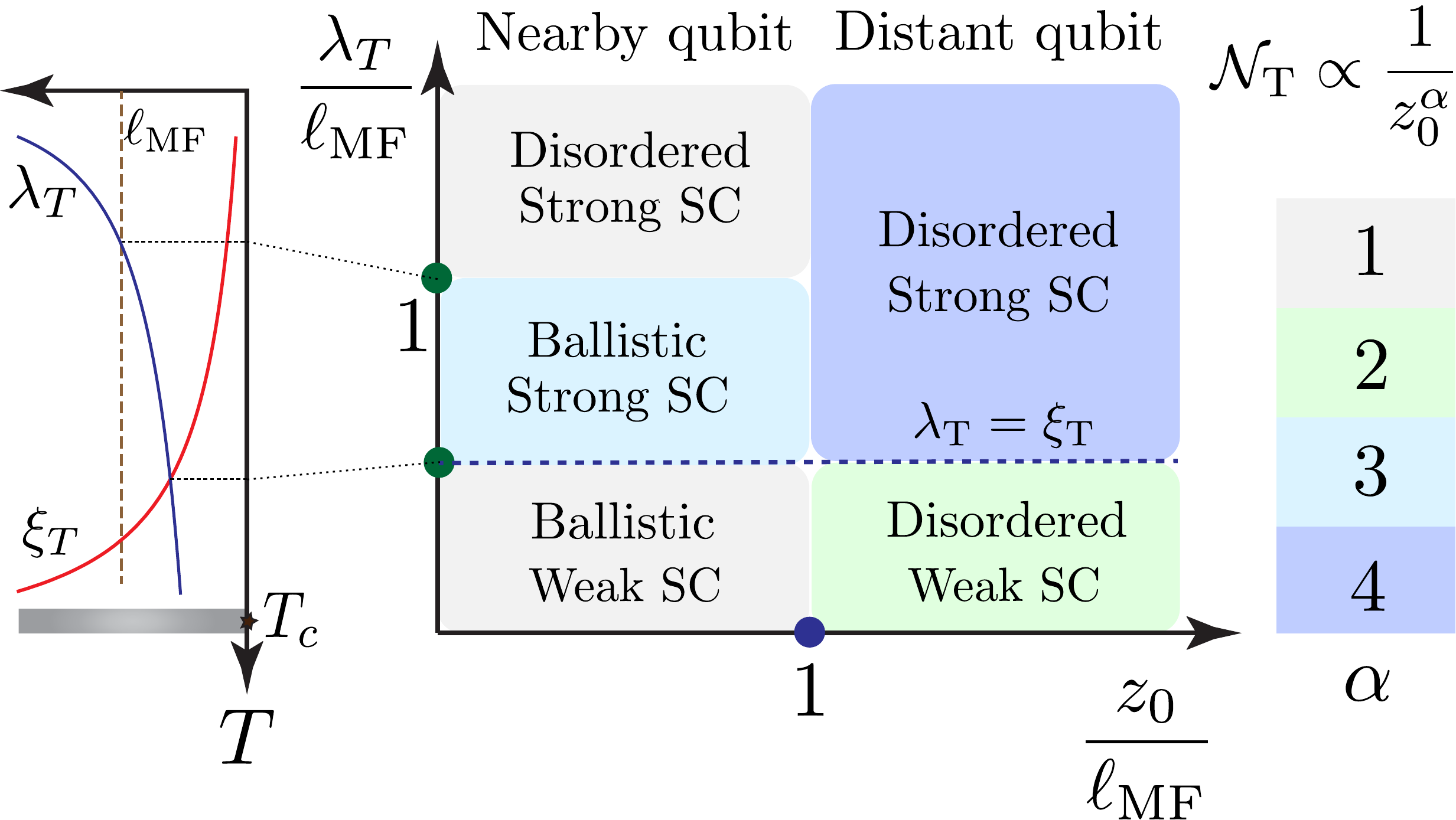}
    \caption{Distance-scaling of noise in different physical regimes in a s-wave superconductor, determined by the ratio of the thermal wavelength $\lambda_{T}$ or qubit-sample distance $z_0$ to the mean-free path $\ell_{\rm MF}$.}
    \label{fig:DistanceScaling}
\end{figure}

\emph{Transverse noise---} In the transverse sector, the noise is given in the experimentally relevant limit $
\Omega \ll k_B T$ by~\cite{Shubhayu2,Agarwal2017} (assuming in-plane rotational symmetry):
\beq
\mathcal{N}_{\rm T}(\Omega) \approx \frac{\mu_0 k_B T}{16 \pi z_0^3 \Omega}\int\limits_0^\infty dx \, x^2 e^{-x} \text{Im}\Big\{ r_s\Big(\frac{x}{2z_0 },\Omega\Big)  \Big\}\label{eqn:NV},
\eeq
where $r_s(\q,\Omega) = -\left(1 + \frac{2i q}{\mu_0 \Omega \, \sigma^{\rm T}(\q,\Omega)}\right)^{-1} $ is determined by the transverse conductivity $\sigma^{\rm T}(\q,\Omega)$.
Therefore, the computation of noise reduces to evaluating $\sigma^{\rm T}(\q,\Omega)$ for the 2D sample of interest. 
For the sake of simplicity and physical transparency, we treat the superconductor within the two-fluid model~\cite{Tinkham}. 
The transverse conductivity consists of a dissipative part $\sigma_n^{\rm T}$ due to the normal fluid (quasiparticle excitations) and a reactive part due to transverse superflow, which can be computed using London's equation $\j_{\rm T,s} = - \Lambda \A_{\rm T} = - \left(\frac{\Lambda}{i \Omega} \right) \E_{\rm T}$: 
\beq
\sigma^{\rm T}(\q,\Omega) = \sigma^{\rm T}_n(\q,\Omega) - \frac{\Lambda}{i \Omega} .
\eeq
Within the phenomenological Ginzburg-Landau approach, $\Lambda \propto |\Delta(T)|^2 \propto T_c - T$, $\Delta(T)$ 
being the superconducting gap. Therefore, as $T$ approaches $T_c$,  
the reactive part can be neglected, and the conductivity is dominated by the quasiparticle contribution:
\beq
\mathcal{N}_{\rm T}(\Omega) \approx  \frac{\mu_0^2 k_B T}{16 \pi z^2_{0}}  \text{Re}\left\{\sigma_n^{\rm T}\Big(\frac{1}{2z_0},\Omega\Big)\right\}
\label{eq:Nz1}.
\eeq
In contrast, deep in the superconducting phase ($k_B T \ll \Delta(T)$, indicated by ``strong SC" in Fig.~\ref{fig:DistanceScaling}) at small frequency $\Omega$, the quasiparticle contribution is suppressed and conductivity is dominated by superflow\footnote{Strictly speaking, this formula is valid for $z_0 $ being greater than the ``Pearl length" $1/\mu_0\Lambda$ \cite{pearl1964current}. See Ref.~\cite{Shubhayu2} for details.}:
\beq
\mathcal{N}_{\rm T}(\Omega) &\approx & \frac{3 k_B T}{8 \pi z_0^4\Lambda^2}\text{Re}\left\{\sigma_n^{\rm T} \Big(\frac{3}{2z_0},\Omega\Big) \right\}
\label{eq:Nz2}.
\eeq
We see that in both limits, the essential signatures of the properties of the superconducting sample can be gleaned from Re$\{ \sigma^{\rm T}_n(q,\Omega)\}$, which we discuss next. 

To calculate $\sigma^{\rm T}_n$ for the \textit{normal fluid}, we resort to a microscopic description of the current-carrying quasiparticle excitations via the mean-field BCS Hamiltonian $H_{\rm BCS}$. 
For singlet superconductors, $H_{\rm BCS}$ is given in terms of the Nambu spinor $\Psi_{\k} = (c_{\k, \uparrow}, c^\dagger_{-\k, \downarrow})^T$ (where $c_{\k, \sigma}$ are electron annihilation operators), electron dispersion $\xi_{\k} = \frac{k^2}{2m} - \mu$, and gap-function $\Delta_\k$:
\beq
H_{\rm BCS} = \sum_{\k} \Psi^\dagger_\k h_\k \Psi_{\k}, ~ h_\k = \begin{pmatrix} \xi_\k & \Delta_\k \\ \Delta_\k & -\xi_{\k} \end{pmatrix}.
\eeq
Accordingly, the retarded Green's function is given by $G^R(\k, \omega) = ( \omega + i 0^+ + i \Gamma_0 - h_\k)^{-1} $, where $\Gamma_0$ is the isotropic scattering rate of electrons at the Fermi surface due to uncorrelated quenched disorder, and it accounts for the broadening of the quasiparticle spectrum~\cite{DurstLee}. 
Denoting the transverse current density by $j_{\rm T}(\q) = (\hat{z} \times \hat{\q}) \cdot  \j(\q)$, the conductivity can be evaluated within the standard linear response formalism~\cite{altland2010condensed}, using the retarded transverse current-current correlation function $ \Pi^R_{\rm T}(\q, t) = -i\theta(t) \langle [ j_{\rm T}(\q, t), j_{\rm T}(-\q, 0)] \rangle/{\cal A}$:
\beq 
\label{eq:sigmaN}
&& \text{Re}[\sigma^{\rm T}_n(\q,\Omega)] =  -\frac{\text{Im}[\Pi^R_{\rm T}(\q, \Omega)]}{\Omega} \\  
&& \xrightarrow{\beta \Omega \ll 1}  \frac{e^2}{2\pi}  \int_{\k,\omega} \; v_{\rm T}^2 \Tr[A(\k_-,\omega) A(\k_+, \omega_+)] \left( -\frac{\partial n_F}{\partial \omega} \right), \nonumber
\eeq
where $\bm{v}_{\rm T}(\hat{k}) = v_F (\hat{q} \times \hat{k})$ is the transverse component of Fermi velocity, $\k_\pm = \k \pm \q/2$, $\omega_+ = \omega + \Omega$, $A(\k, \Omega) = - \text{Im}[G^R(\k, \Omega)]/\pi$ is the spectral function, ${\cal A}$ is the area of the sample, and $n_F(\omega) = [\exp(\beta\omega) + 1]^{-1}$ is the Fermi function ($\beta = 1/k_B T$).
A key point to note is that as a consequence of momentum and energy selectivity, the qubit probe is most sensitive to $\sigma^{\rm T}_n(q \sim 1/z_0, \Omega \approx 0)$, since for realistic experimental scenarios the relevant dimensionless variable $\Omega/(v_F q) \sim \Omega z_0/v_F \ll 1$. 
This is a completely distinct order of limits from the usual computation of dc conductivity, where one first takes $\q \to 0$ and then $\Omega \to 0$. 
As a consequence, our qubit probe opens the door to studying the finite-momentum low-frequency transverse conductivity, which yields information about intrinsic behavior that is typically inaccessible otherwise.

The behavior of $\sigma^{\rm T}_n(\q,\Omega)$ is generally quite complex, and is determined by the rich interplay of several different length-scales:
(i) the quasiparticle mean-free path $\ell_{\rm MF} = v_F/(2 \Gamma_0)$, (ii) the qubit-sample distance $z_0$, (iii) the quasiparticle thermal wavelength $\lambda_T \equiv v_F/k_B T$, and (iv) the superconducting coherence length $\xi_T \equiv v_F/\Delta(T)$. Varying these parameters by tuning $z_0$ and $T$ results in various crossovers in the behavior of $\sigma_n^{\rm T}(q\sim 1/z_0,0)$, which can then be used to infer important properties of the superconducting state. Before getting into the details of these crossovers, we highlight the essential gross features: Above $T_c$, $\sigma_n^{\rm T}(\q,0)$ is approximately independent of $T$, resulting in $\mathcal{N}_{\rm T} \propto T$. Upon cooling through $T_c$, a significant suppression of noise occurs, providing a clear indication of the superconducting transition. In an s-wave superconductor, quasiparticle excitations are suppressed  by the gap, so that the noise decreases exponentially as $e^{-\Delta(T)/k_B T}$. In a d-wave supercondcutor, the presence of nodal quasiparticles yields a $T^2$ behavior. Distinct temperature scalings of the noise therefore provide a way to distinguish superconducting order parameters.

In addition to temperature dependence, the scaling of noise with distance is different for nodal vs non-nodal superconductors: In the temperature regime $k_B T \lesssim \Gamma_0$, we find $\mathcal{N}_{\rm T} \propto 1/z_0$ in the s-wave case, while $\mathcal{N}_{\rm T} \propto 1/z_0^2$ for the d-wave one. We will now elaborate further on the behavior $\sigma_n^{\rm T}(q\sim 1/z_0,0)$, appealing to physical arguments to demonstrate how the distance-scaling of noise can be used to distinguish different transport regimes. All our conclusions are verified by explicit analytic and numerical evaluation of noise across arbitrary parameter regimes~\cite{Shubhayu2}.

We first consider 2D s-wave superconductors in the clean limit, where the thermal wavelength $\lambda_T$ is much smaller than the mean-free path $\ell_{\rm MF}$, or equivalently $\Gamma_0 \lesssim k_B T$.
In this limit, the spectral function $A(\k,\omega)$ is not sufficiently smoothed out by disorder on the scale of $|\omega| \lesssim k_B T$, the frequency range where $n_F^\prime(\omega)$ is significant.
Hence, it is useful to think of the conductivity as arising from sharp transitions of quasiparticles near the Fermi surface across an energy shell of width $\Omega$ and a momentum transfer $\q$. 
The phase space of such excitations is constrained by the geometry of the Fermi surface and grows as $1/q$.  
Accordingly, at low temperatures $k_B T \ll \Delta(T)$, the conductivity scales as $\sigma_n^{\rm T}(\q, 0) \propto e^{- \beta \Delta}(q T \xi_T)^{-1}$.
Considering additional suppression due to the super-flow in this regime, Eq.~\eqref{eq:Nz2} predicts $\mathcal{N}_{\rm T} \propto 1/z_0^{3}$.
On increasing $T$, as the Cooper pairs get loosely bound and $\xi_T$ crosses the thermal length $\lambda_T$, we recover the conductivity of a ballistic Fermi liquid by simply replacing $\xi_T$ by $\lambda_T$ in the above expression, i.e, $\sigma_n^{\rm T}(\q, 0) \propto (q T \lambda_T)^{-1} = (q v_F)^{-1}$ and $\mathcal{N}_{\rm T} \propto 1/z_0$ \cite{Khoo}.

In the dirty limit, $\Gamma_0 \gtrsim k_B T$, the spectral function is smooth on the scale of $\omega \lesssim k_B T$, and we can replace the Fermi function derivative by a delta function at $\omega = 0$. 
In this disordered superconductor, the conductivity is determined by the product of spectral functions near the Fermi surface (see Eq.~\eqref{eq:sigmaN}).
If we lower the sample-probe distance to probe large momenta, i.e, $q\ell_{\rm MF}$ and $q\xi_T$ are both large (but $q \ll k_F$), then the product $A(\k_+,0)A(\k_-,0) \sim 1/(v_F q)^4$ for $|\k| \approx k_F$.
The dominant contribution to transverse conductivity comes from an annular region of width $q$ and circumference $2\pi k_F$ around the circular Fermi surface.
Accordingly, the conductivity scales as $2 \pi k_F \times q \times 1/q^4 \sim k_F/q^3$, and Eq.~\eqref{eq:Nz2} predicts that the noise decreases as $1/z_0$.
On approaching the opposite limit $z_0 \gg \ell_{\rm MF}$, the probe is sensitive to multiple scattering events, resulting in the usual metallic behavior, where $\sigma_n^{\rm T}(\q, 0)$ is independent of momentum and $\mathcal{N}_{\rm T} \propto 1/z_0^{4}$. The noise scalings in different regimes are summarized in Fig.~\ref{fig:DistanceScaling}, along with the behavior in the regime $\Delta(T) \lesssim k_B T$, which we refer to as the ``weak superconductor" (weak SC). In the latter regime, the scaling is the same as that in the metallic phase above $T_c$.

Next, we consider $d$-wave superconductors with the gap function $\Delta_\k(T) = \Delta_0(T)(k_x^2 - k_y^2)$. 
We find that the distinct temperature-scaling of the noise carries signatures of gapless Dirac quasiparticles at the nodes $k = k_F, k_x = \pm k_y$.
In the clean (ballistic) limit, at low $k_B T \ll \Delta_0(T)$, the noise scales as $T^2$ due to the power-law low-energy density of states at the Dirac cones.
As $T \to T_c^{-}$, the noise increases steeply as $v_\Delta^{-1}\ln(v_F/v_\Delta)$, where the gap velocity $v_\Delta = |\partial_\k \Delta_\k|_{\rm node}$ decreases as $|T_c - T|^{1/2}$ within mean-field theory. 
This logarithmic correction owes its origin to the anisotropy between the Fermi velocity $v_F$ and the gap velocity $v_\Delta$ --- the dominant contribution comes from the tips of anisotropic Dirac cones with large density of states.
In the dirty limit, the disorder induced finite density of states at zero energy leads to a linear in $T$ scaling of the noise, as the conductivity $\sigma_n^{\rm T}(\q,\Omega = 0)$ approaches a constant non-singular value of $e^2v_F/\pi^2 v_\Delta$ at small $q$ \cite{DurstLee}. 

The physics of distance scalings remains roughly the same as for the s-wave case, with the exception of dirty superconductors at sample-probe distances smaller than the mean-free path. 
The available phase-space for excitations constitute a patch of area $q^2$ around each nodal points, implying $\sigma_n^{\rm T}(\q,0)\propto q^2 \times 1/q^{4} = 1/q^{2}$.
In this limit, the noise scales as $1/z_0^2$, up to logarithmic corrections, and can be used to differentiate nodal and non-nodal superconducting gap functions.

\begin{figure}
    \centering
    \includegraphics[width=0.35\textwidth]{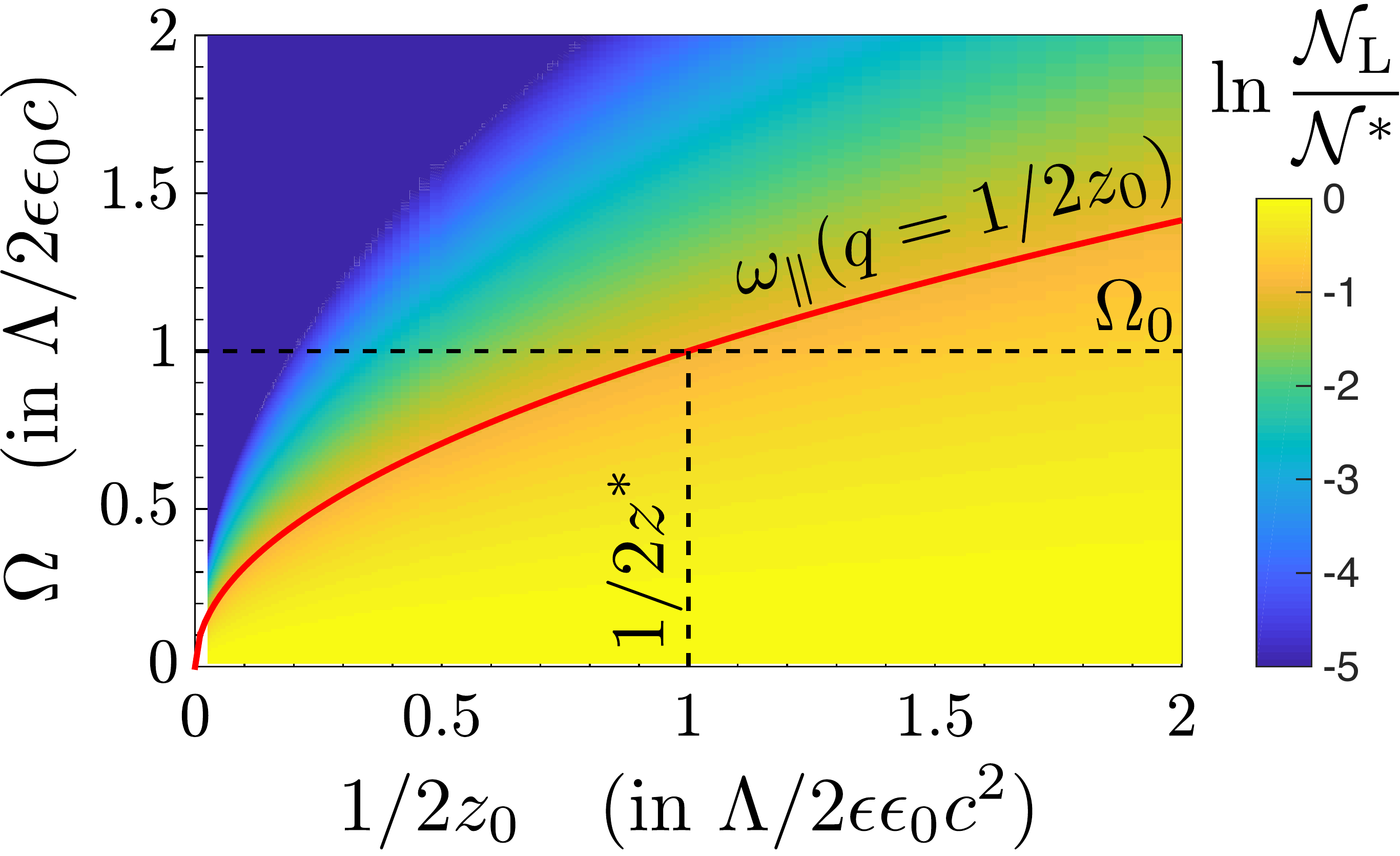}
    \caption{ Plasmon contribution to the low-temperature noise. Shown is $\ln[{\cal N}_{\rm L}(\Omega,z_0)/{\cal N}^*(\Omega)]$ as a function of $\Omega$ and $z_{0}$. For large $z_0\gg z^*(\Omega) \equiv 4 \epsilon \epsilon_0 \Lambda/\Omega^2$ (such that $\Omega = \omega_\parallel(1/2z^*)$), the noise is suppressed and it quickly saturates upon crossing the plasmon dispersion (solid red line), i.e. for $z_{0}\lesssim z^*(\Omega)$.}
    \label{fig:Longitudinal}
\end{figure}

\emph{Noise from longitudinal collective modes---} As remarked, at low temperatures, the quasiparticle noise in the transverse sector is heavily suppressed, and longitudinal fluctuations can dominate. 
Longitudinal noise is given by~\cite{Shubhayu2}:
\begin{equation}
\mathcal{N}_{\rm L}(\Omega) = \frac{k_B T \epsilon \mu_0 \Omega }{16 \pi c^2 z_0} \int_0^\infty dx \, e^{-x} \text{Im} \left\{ r_p\left( \frac{x}{2 z_0}, \Omega \right) \right\},
\end{equation}
where $r_p(\q, \Omega)= \left(  1 + \frac{2 \epsilon \epsilon_0 \Omega}{i q \sigma^{\rm L}(\q,\Omega)} \right)^{-1}$ and $\epsilon$ is the high-frequency dielectric constant of the encapsulating medium.
The longitudinal conductivity $\sigma^{\rm L}(\q,\Omega)$ can be evaluated by considering longitudinal current fluctuations via Maxwell's equations and time-dependent Ginburg-Landau theory for the superconducting order parameter \cite{Tinkham}.
For a 2D sample, such an analysis yields a collective plasmon mode with a gapless dispersion: $\omega_\parallel (q) \approx  \sqrt{q\Lambda/(2\epsilon\epsilon_0)}$ \cite{Shubhayu2}.
This collective excitation represents an {\it evanescent} wave that couples three-dimensional fluctuations of the electromagnetic field to two-dimensional quasiparticle currents and order parameter fluctuations.  
Interestingly, we find that the reflection coefficient $r_p$, evaluated for evanescent waves, exhibits a sharp peak at the resonance frequency $\Omega = \omega_\parallel(q)$.
Therefore, the longitudinal noise can become resonantly enhanced when the probe distance $z_0$ is decreased to cross the longitudinal spectrum (one could also vary the probe frequency $\Omega$ at fixed $z_0$):
\beq
\mathcal{N}_{\rm L}(\Omega) \approx  \mathcal{N}^*(\Omega)\exp\left(-\frac{4 \epsilon \epsilon_0 z_0 \Omega^2}{\Lambda}\right), \label{eqn:N_x}
\eeq
where $\mathcal{N}^*(\Omega) =  k_B T \epsilon^2 \Omega^3/(4 \Lambda c^4)$.
We conclude that although $\sigma_n^{\rm L}$ can be vanishingly small at low temperatures due to superconductivity, the noise still gets a substantial contribution from the longitudinal collective modes.
In particular,  $\mathcal{N}_{\rm L}(\Omega)/\mathcal{N}^*(\Omega)$ is suppressed for large sample-probe distances, whereas it saturates to a finite value for $z_0\lesssim z^*(\Omega)$, defined by $\Omega = \omega_\parallel(1/2z^*)$, i.e. upon crossing the plasmon branch, as illustrated in Fig.~\ref{fig:Longitudinal}. We also note that, in bilayers, interlayer Josephson plasmons lead to an additional resonance in the longitudinal noise (see Ref.~\cite{Shubhayu2} for further discussion). 

\emph{Experimental feasibility---} To be concrete, let us consider NV centers in diamond \cite{hong2013nanoscale,nvsinglespin,nvreview,Casola,Stano2013}. 
The intrinsic level splitting is $\Omega/2\pi = 2.87$ GHz ($\approx 0.1 K$), the smallest energy scale in the problem.
This justifies taking $\Omega \to 0$ in our calculations. 
For magic angle moir\'{e} graphene, $T_c \approx 3-4\,K$ \cite{cao2018,park2021,hao2021}.
Assuming sufficient disorder (due to local strain \cite{Kazmierczak_2021}), we approximate $\sigma^{\rm T}(q) \approx \sigma(q=0) \approx 10^{-2}\,S$ in the metallic state just above $T_c$. 
For $z_0 = 5$\,nm, we can therefore estimate $\mathcal{N}_{\rm T} \approx \frac{k_B T\mu_0^2 \sigma^{\rm T}_{n}}{16 \pi z_0^2} \approx 120$\,pT$^2$/Hz.
Accordingly, the depolarization rate of the qubit is given by \cite{Agarwal2017} $1/T_1 = \sqrt{S(S+1)} g^2 \mu_B^2 \mathcal{N}_{\rm T}/(2\hbar^2)  \approx 3.2$\,s$^{-1}$ ($S=1$ for NV).
The lifetime in the presence of a sample is therefore much shorter than the intrinsic lifetime of an isolated NV center, which can exceed $10^2$ s below 10 K \cite{Jarmola,Guillebon}.
Further, within mean-field theory $\Delta(T) \approx 3 k_B T_c \sqrt{1 - T/T_c}$ \cite{Tinkham}, implying that on decreasing $T$ from $T_c$ to $T_c/2$, $1/T_1$ drops roughly by a factor of $e^{\beta \Delta(T)} \approx 70$, making the change easily detectable. Comparable parameters apply to TMDs, which exhibit a variety of interesting superconducting behaviors \cite{Manzeli2017}, such as ``Ising superconductivity".
The foregoing estimates show that the detection of the ordering transition is realistic within the existing experimental tools. However, to detect the longitudinal collective modes, one might need to use high-$\epsilon$ materials, such as SrTiO$_3$ with $\epsilon\approx 10000$~\cite{Veyrat781}. 
Another way to enhance this signal is to increase the probe frequency $\Omega$ via the application of an in-plane magnetic field~\cite{Shubhayu2}.

\emph{Discussion---}
Our work establishes a complimentary route to detect and study 2D superconductivity via noise magnetometry and offers a window into the elusive Coulomb-interaction-dominated longitudinal collective modes in superconductors. 
The ability to probe across a wide range of length scales allows our technique to investigate inhomogeneous materials and to potentially understand how superconductivity differs across samples.
Our calculations are valid away from the critical regime, where critical current fluctuations might lead to an appreciable enhancement of noise, facilitating easier experimental detection.
This exciting direction deserves further investigation, though we note that this critical temperature window is typically relatively narrow in superconductors.

While our qubit probe is also sensitive to noise arising from spin fluctuations, spin-noise in singlet superconductors is suppressed relative to charge noise by an additional factor of $1/(k_F z_0)^2$ \cite{Agarwal2017} in the nearly metallic regime and by $(\mu_0 \mu_B \Lambda/e v_F)^2$ deep in the superconducting phase \cite{Shubhayu2}. Further, in contrast to NMR, there is no anomalous enhancement of spin-noise (relative to current-noise), as the non-local response leads to smearing out the singular density of states that is responsible for the Hebel-Slichter peak~\cite{Shubhayu2}.
However, in superconductors with spontaneously broken spin-rotation symmetry (in materials with negligible spin-orbit coupling), spin-noise may receive resonant enhancement from spin-waves \cite{CRD18,Joaquin_magnon_18}.

Previous works have developed theoretical underpinnings of probing different regimes of electron transport in metallic systems, as well as magnetic phases and their phase transitions in insulators, using quantum impurity probes \cite{Agarwal2017,Joaquin18,CRD18}. 
On the experimental front, noise measurements using single-spin qubits have recently been used to detect a plethora of novel quantum phenomena, including non-local conductivity in metallic silver \cite{Kolkowitz}, magnetic transitions in metallic Gd \cite{Satcher}, chiral magnons \cite{Rustagi}, spin-diffusion in antiferromagnets \cite{Du} and electron-phonon instabilities in graphene \cite{AndersenDwyer}. 
With the discovery of superconductivity in several new 2D materials, the qubit probe appears ideally suited to unravel their properties. 
The non-invasive nature of the qubit, with fully optical initialization and readout capabilities, adds to its appeal.
Finally, we note that current fluctuations in the sample also provide electric field noise at the qubit location \cite{PhysRevLett.118.197201,Sahay2021} and alternative qubits
~\cite{gottscholl2019room,Tetienne} are more susceptible to this kind of noise.
Their response can be studied by a generalization of the theoretical framework developed here and is left for future work.

\emph{Acknowledgments---} We thank T. Andersen, B. Dwyer, S. Hsieh, S. Kolkowitz, V. Manucharian, J. F. Rodriguez-Nieva, E. Urbach, R. Xue, A. Yacoby and C. Zu for helpful conversations.
S.C. was supported by the ARO through the Anyon Bridge MURI program (Grant No. W911NF-17-1-0323) via M. P. Zaletel, and the  U.S. DOE, Office of Science, Office of Advanced Scientific Computing Research, under the Accelerated Research in Quantum Computing (ARQC) program via N.Y. Yao.
P.E.D, I.E., and E.D. were supported by Harvard-MIT CUA, AFOSR-MURI: Photonic Quantum Matter Award No. FA95501610323, Harvard Quantum Initiative, and AFOSR Grant No. FA9550-21-1-0216. N.Y.Y. acknowledges support from U.S. DOE, Office of Science, Office of Advanced Scientific Computing Research Quantum Testbed Program.
Support from the Gordon and Betty Moore Foundation is also gratefully acknowledged.

\bibliography{Sc_NV}

\begin{thebibliography}{48}%
\makeatletter
\providecommand \@ifxundefined [1]{%
 \@ifx{#1\undefined}
}%
\providecommand \@ifnum [1]{%
 \ifnum #1\expandafter \@firstoftwo
 \else \expandafter \@secondoftwo
 \fi
}%
\providecommand \@ifx [1]{%
 \ifx #1\expandafter \@firstoftwo
 \else \expandafter \@secondoftwo
 \fi
}%
\providecommand \natexlab [1]{#1}%
\providecommand \enquote  [1]{``#1''}%
\providecommand \bibnamefont  [1]{#1}%
\providecommand \bibfnamefont [1]{#1}%
\providecommand \citenamefont [1]{#1}%
\providecommand \href@noop [0]{\@secondoftwo}%
\providecommand \href [0]{\begingroup \@sanitize@url \@href}%
\providecommand \@href[1]{\@@startlink{#1}\@@href}%
\providecommand \@@href[1]{\endgroup#1\@@endlink}%
\providecommand \@sanitize@url [0]{\catcode `\\12\catcode `\$12\catcode
  `\&12\catcode `\#12\catcode `\^12\catcode `\_12\catcode `\%12\relax}%
\providecommand \@@startlink[1]{}%
\providecommand \@@endlink[0]{}%
\providecommand \url  [0]{\begingroup\@sanitize@url \@url }%
\providecommand \@url [1]{\endgroup\@href {#1}{\urlprefix }}%
\providecommand \urlprefix  [0]{URL }%
\providecommand \Eprint [0]{\href }%
\providecommand \doibase [0]{http://dx.doi.org/}%
\providecommand \selectlanguage [0]{\@gobble}%
\providecommand \bibinfo  [0]{\@secondoftwo}%
\providecommand \bibfield  [0]{\@secondoftwo}%
\providecommand \translation [1]{[#1]}%
\providecommand \BibitemOpen [0]{}%
\providecommand \bibitemStop [0]{}%
\providecommand \bibitemNoStop [0]{.\EOS\space}%
\providecommand \EOS [0]{\spacefactor3000\relax}%
\providecommand \BibitemShut  [1]{\csname bibitem#1\endcsname}%
\let\auto@bib@innerbib\@empty
%</preamble>
\bibitem [{\citenamefont {Tinkham}(2004)}]{Tinkham}%
  \BibitemOpen
  \bibfield  {author} {\bibinfo {author} {\bibfnamefont {Michael}\ \bibnamefont
  {Tinkham}},\ }\href@noop {} {\emph {\bibinfo {title} {Introduction to
  superconductivity}}}\ (\bibinfo  {publisher} {Courier Corporation},\ \bibinfo
  {year} {2004})\BibitemShut {NoStop}%
\bibitem [{\citenamefont {Cao}\ \emph {et~al.}(2018)\citenamefont {Cao},
  \citenamefont {Fatemi}, \citenamefont {Fang}, \citenamefont {Watanabe},
  \citenamefont {Taniguchi}, \citenamefont {Kaxiras},\ and\ \citenamefont
  {Jarillo-Herrero}}]{cao2018}%
  \BibitemOpen
  \bibfield  {author} {\bibinfo {author} {\bibfnamefont {Yuan}\ \bibnamefont
  {Cao}}, \bibinfo {author} {\bibfnamefont {Valla}\ \bibnamefont {Fatemi}},
  \bibinfo {author} {\bibfnamefont {Shiang}\ \bibnamefont {Fang}}, \bibinfo
  {author} {\bibfnamefont {Kenji}\ \bibnamefont {Watanabe}}, \bibinfo {author}
  {\bibfnamefont {Takashi}\ \bibnamefont {Taniguchi}}, \bibinfo {author}
  {\bibfnamefont {Efthimios}\ \bibnamefont {Kaxiras}}, \ and\ \bibinfo {author}
  {\bibfnamefont {Pablo}\ \bibnamefont {Jarillo-Herrero}},\ }\bibfield  {title}
  {\enquote {\bibinfo {title} {Unconventional superconductivity in magic-angle
  graphene superlattices},}\ }\href@noop {} {\bibfield  {journal} {\bibinfo
  {journal} {Nature}\ }\textbf {\bibinfo {volume} {556}},\ \bibinfo {pages}
  {43--50} (\bibinfo {year} {2018})}\BibitemShut {NoStop}%
\bibitem [{\citenamefont {Lu}\ \emph {et~al.}(2019)\citenamefont {Lu},
  \citenamefont {Stepanov}, \citenamefont {Yang}, \citenamefont {Xie},
  \citenamefont {Aamir}, \citenamefont {Das}, \citenamefont {Urgell},
  \citenamefont {Watanabe}, \citenamefont {Taniguchi}, \citenamefont {Zhang}
  \emph {et~al.}}]{lu2019}%
  \BibitemOpen
  \bibfield  {author} {\bibinfo {author} {\bibfnamefont {Xiaobo}\ \bibnamefont
  {Lu}}, \bibinfo {author} {\bibfnamefont {Petr}\ \bibnamefont {Stepanov}},
  \bibinfo {author} {\bibfnamefont {Wei}\ \bibnamefont {Yang}}, \bibinfo
  {author} {\bibfnamefont {Ming}\ \bibnamefont {Xie}}, \bibinfo {author}
  {\bibfnamefont {Mohammed~Ali}\ \bibnamefont {Aamir}}, \bibinfo {author}
  {\bibfnamefont {Ipsita}\ \bibnamefont {Das}}, \bibinfo {author}
  {\bibfnamefont {Carles}\ \bibnamefont {Urgell}}, \bibinfo {author}
  {\bibfnamefont {Kenji}\ \bibnamefont {Watanabe}}, \bibinfo {author}
  {\bibfnamefont {Takashi}\ \bibnamefont {Taniguchi}}, \bibinfo {author}
  {\bibfnamefont {Guangyu}\ \bibnamefont {Zhang}},  \emph {et~al.},\ }\bibfield
   {title} {\enquote {\bibinfo {title} {Superconductors, orbital magnets and
  correlated states in magic-angle bilayer graphene},}\ }\href@noop {}
  {\bibfield  {journal} {\bibinfo  {journal} {Nature}\ }\textbf {\bibinfo
  {volume} {574}},\ \bibinfo {pages} {653--657} (\bibinfo {year}
  {2019})}\BibitemShut {NoStop}%
\bibitem [{\citenamefont {Yankowitz}\ \emph {et~al.}(2019)\citenamefont
  {Yankowitz}, \citenamefont {Chen}, \citenamefont {Polshyn}, \citenamefont
  {Zhang}, \citenamefont {Watanabe}, \citenamefont {Taniguchi}, \citenamefont
  {Graf}, \citenamefont {Young},\ and\ \citenamefont {Dean}}]{yankowitz2019}%
  \BibitemOpen
  \bibfield  {author} {\bibinfo {author} {\bibfnamefont {Matthew}\ \bibnamefont
  {Yankowitz}}, \bibinfo {author} {\bibfnamefont {Shaowen}\ \bibnamefont
  {Chen}}, \bibinfo {author} {\bibfnamefont {Hryhoriy}\ \bibnamefont
  {Polshyn}}, \bibinfo {author} {\bibfnamefont {Yuxuan}\ \bibnamefont {Zhang}},
  \bibinfo {author} {\bibfnamefont {K}~\bibnamefont {Watanabe}}, \bibinfo
  {author} {\bibfnamefont {T}~\bibnamefont {Taniguchi}}, \bibinfo {author}
  {\bibfnamefont {David}\ \bibnamefont {Graf}}, \bibinfo {author}
  {\bibfnamefont {Andrea~F}\ \bibnamefont {Young}}, \ and\ \bibinfo {author}
  {\bibfnamefont {Cory~R}\ \bibnamefont {Dean}},\ }\bibfield  {title} {\enquote
  {\bibinfo {title} {Tuning superconductivity in twisted bilayer graphene},}\
  }\href@noop {} {\bibfield  {journal} {\bibinfo  {journal} {Science}\ }\textbf
  {\bibinfo {volume} {363}},\ \bibinfo {pages} {1059--1064} (\bibinfo {year}
  {2019})}\BibitemShut {NoStop}%
\bibitem [{\citenamefont {{Park}}\ \emph {et~al.}(2021)\citenamefont {{Park}},
  \citenamefont {{Cao}}, \citenamefont {{Watanabe}}, \citenamefont
  {{Taniguchi}},\ and\ \citenamefont {{Jarillo-Herrero}}}]{park2021}%
  \BibitemOpen
  \bibfield  {author} {\bibinfo {author} {\bibfnamefont {Jeong~Min}\
  \bibnamefont {{Park}}}, \bibinfo {author} {\bibfnamefont {Yuan}\ \bibnamefont
  {{Cao}}}, \bibinfo {author} {\bibfnamefont {Kenji}\ \bibnamefont
  {{Watanabe}}}, \bibinfo {author} {\bibfnamefont {Takashi}\ \bibnamefont
  {{Taniguchi}}}, \ and\ \bibinfo {author} {\bibfnamefont {Pablo}\ \bibnamefont
  {{Jarillo-Herrero}}},\ }\bibfield  {title} {\enquote {\bibinfo {title}
  {{Tunable strongly coupled superconductivity in magic-angle twisted trilayer
  graphene}},}\ }\href {\doibase 10.1038/s41586-021-03192-0} {\bibfield
  {journal} {\bibinfo  {journal} {\nat}\ }\textbf {\bibinfo {volume} {590}},\
  \bibinfo {pages} {249--255} (\bibinfo {year} {2021})}\BibitemShut {NoStop}%
\bibitem [{\citenamefont {Hao}\ \emph {et~al.}(2021)\citenamefont {Hao},
  \citenamefont {Zimmerman}, \citenamefont {Ledwith}, \citenamefont {Khalaf},
  \citenamefont {Najafabadi}, \citenamefont {Watanabe}, \citenamefont
  {Taniguchi}, \citenamefont {Vishwanath},\ and\ \citenamefont
  {Kim}}]{hao2021}%
  \BibitemOpen
  \bibfield  {author} {\bibinfo {author} {\bibfnamefont {Zeyu}\ \bibnamefont
  {Hao}}, \bibinfo {author} {\bibfnamefont {A.~M.}\ \bibnamefont {Zimmerman}},
  \bibinfo {author} {\bibfnamefont {Patrick}\ \bibnamefont {Ledwith}}, \bibinfo
  {author} {\bibfnamefont {Eslam}\ \bibnamefont {Khalaf}}, \bibinfo {author}
  {\bibfnamefont {Danial~Haie}\ \bibnamefont {Najafabadi}}, \bibinfo {author}
  {\bibfnamefont {Kenji}\ \bibnamefont {Watanabe}}, \bibinfo {author}
  {\bibfnamefont {Takashi}\ \bibnamefont {Taniguchi}}, \bibinfo {author}
  {\bibfnamefont {Ashvin}\ \bibnamefont {Vishwanath}}, \ and\ \bibinfo {author}
  {\bibfnamefont {Philip}\ \bibnamefont {Kim}},\ }\bibfield  {title} {\enquote
  {\bibinfo {title} {Electric field–tunable superconductivity in
  alternating-twist magic-angle trilayer graphene},}\ }\href {\doibase
  10.1126/science.abg0399} {\bibfield  {journal} {\bibinfo  {journal}
  {Science}\ }\textbf {\bibinfo {volume} {371}},\ \bibinfo {pages}
  {1133–1138} (\bibinfo {year} {2021})}\BibitemShut {NoStop}%
\bibitem [{\citenamefont {{Shi}}\ \emph {et~al.}(2015)\citenamefont {{Shi}},
  \citenamefont {{Ye}}, \citenamefont {{Zhang}}, \citenamefont {{Suzuki}},
  \citenamefont {{Yoshida}}, \citenamefont {{Miyazaki}}, \citenamefont
  {{Inoue}}, \citenamefont {{Saito}},\ and\ \citenamefont {{Iwasa}}}]{shi2015}%
  \BibitemOpen
  \bibfield  {author} {\bibinfo {author} {\bibfnamefont {Wu}~\bibnamefont
  {{Shi}}}, \bibinfo {author} {\bibfnamefont {Jianting}\ \bibnamefont {{Ye}}},
  \bibinfo {author} {\bibfnamefont {Yijin}\ \bibnamefont {{Zhang}}}, \bibinfo
  {author} {\bibfnamefont {Ryuji}\ \bibnamefont {{Suzuki}}}, \bibinfo {author}
  {\bibfnamefont {Masaro}\ \bibnamefont {{Yoshida}}}, \bibinfo {author}
  {\bibfnamefont {Jun}\ \bibnamefont {{Miyazaki}}}, \bibinfo {author}
  {\bibfnamefont {Naoko}\ \bibnamefont {{Inoue}}}, \bibinfo {author}
  {\bibfnamefont {Yu}~\bibnamefont {{Saito}}}, \ and\ \bibinfo {author}
  {\bibfnamefont {Yoshihiro}\ \bibnamefont {{Iwasa}}},\ }\bibfield  {title}
  {\enquote {\bibinfo {title} {{Superconductivity Series in Transition Metal
  Dichalcogenides by Ionic Gating}},}\ }\href {\doibase 10.1038/srep12534}
  {\bibfield  {journal} {\bibinfo  {journal} {Scientific Reports}\ }\textbf
  {\bibinfo {volume} {5}},\ \bibinfo {eid} {12534} (\bibinfo {year}
  {2015})}\BibitemShut {NoStop}%
\bibitem [{\citenamefont {{Ge}}\ \emph {et~al.}(2015)\citenamefont {{Ge}},
  \citenamefont {{Liu}}, \citenamefont {{Liu}}, \citenamefont {{Gao}},
  \citenamefont {{Qian}}, \citenamefont {{Xue}}, \citenamefont {{Liu}},\ and\
  \citenamefont {{Jia}}}]{ge2015FeSe}%
  \BibitemOpen
  \bibfield  {author} {\bibinfo {author} {\bibfnamefont {Jian-Feng}\
  \bibnamefont {{Ge}}}, \bibinfo {author} {\bibfnamefont {Zhi-Long}\
  \bibnamefont {{Liu}}}, \bibinfo {author} {\bibfnamefont {Canhua}\
  \bibnamefont {{Liu}}}, \bibinfo {author} {\bibfnamefont {Chun-Lei}\
  \bibnamefont {{Gao}}}, \bibinfo {author} {\bibfnamefont {Dong}\ \bibnamefont
  {{Qian}}}, \bibinfo {author} {\bibfnamefont {Qi-Kun}\ \bibnamefont {{Xue}}},
  \bibinfo {author} {\bibfnamefont {Ying}\ \bibnamefont {{Liu}}}, \ and\
  \bibinfo {author} {\bibfnamefont {Jin-Feng}\ \bibnamefont {{Jia}}},\
  }\bibfield  {title} {\enquote {\bibinfo {title} {{Superconductivity above 100
  K in single-layer FeSe films on doped SrTiO$_{3}$}},}\ }\href {\doibase
  10.1038/nmat4153} {\bibfield  {journal} {\bibinfo  {journal} {Nature
  Materials}\ }\textbf {\bibinfo {volume} {14}},\ \bibinfo {pages} {285--289}
  (\bibinfo {year} {2015})},\ \Eprint {http://arxiv.org/abs/1406.3435}
  {arXiv:1406.3435 [cond-mat.supr-con]} \BibitemShut {NoStop}%
\bibitem [{\citenamefont {{Wang}}\ \emph {et~al.}(2012)\citenamefont {{Wang}},
  \citenamefont {{Li}}, \citenamefont {{Zhang}}, \citenamefont {{Zhang}},
  \citenamefont {{Zhang}}, \citenamefont {{Li}}, \citenamefont {{Ding}},
  \citenamefont {{Ou}}, \citenamefont {{Deng}}, \citenamefont {{Chang}},
  \citenamefont {{Wen}}, \citenamefont {{Song}}, \citenamefont {{He}},
  \citenamefont {{Jia}}, \citenamefont {{Ji}}, \citenamefont {{Wang}},
  \citenamefont {{Wang}}, \citenamefont {{Chen}}, \citenamefont {{Ma}},\ and\
  \citenamefont {{Xue}}}]{Wang2012FeSe}%
  \BibitemOpen
  \bibfield  {author} {\bibinfo {author} {\bibfnamefont {Qing-Yan}\
  \bibnamefont {{Wang}}}, \bibinfo {author} {\bibfnamefont {Zhi}\ \bibnamefont
  {{Li}}}, \bibinfo {author} {\bibfnamefont {Wen-Hao}\ \bibnamefont {{Zhang}}},
  \bibinfo {author} {\bibfnamefont {Zuo-Cheng}\ \bibnamefont {{Zhang}}},
  \bibinfo {author} {\bibfnamefont {Jin-Song}\ \bibnamefont {{Zhang}}},
  \bibinfo {author} {\bibfnamefont {Wei}\ \bibnamefont {{Li}}}, \bibinfo
  {author} {\bibfnamefont {Hao}\ \bibnamefont {{Ding}}}, \bibinfo {author}
  {\bibfnamefont {Yun-Bo}\ \bibnamefont {{Ou}}}, \bibinfo {author}
  {\bibfnamefont {Peng}\ \bibnamefont {{Deng}}}, \bibinfo {author}
  {\bibfnamefont {Kai}\ \bibnamefont {{Chang}}}, \bibinfo {author}
  {\bibfnamefont {Jing}\ \bibnamefont {{Wen}}}, \bibinfo {author}
  {\bibfnamefont {Can-Li}\ \bibnamefont {{Song}}}, \bibinfo {author}
  {\bibfnamefont {Ke}~\bibnamefont {{He}}}, \bibinfo {author} {\bibfnamefont
  {Jin-Feng}\ \bibnamefont {{Jia}}}, \bibinfo {author} {\bibfnamefont
  {Shuai-Hua}\ \bibnamefont {{Ji}}}, \bibinfo {author} {\bibfnamefont {Ya-Yu}\
  \bibnamefont {{Wang}}}, \bibinfo {author} {\bibfnamefont {Li-Li}\
  \bibnamefont {{Wang}}}, \bibinfo {author} {\bibfnamefont {Xi}~\bibnamefont
  {{Chen}}}, \bibinfo {author} {\bibfnamefont {Xu-Cun}\ \bibnamefont {{Ma}}}, \
  and\ \bibinfo {author} {\bibfnamefont {Qi-Kun}\ \bibnamefont {{Xue}}},\
  }\bibfield  {title} {\enquote {\bibinfo {title} {{Interface-Induced
  High-Temperature Superconductivity in Single Unit-Cell FeSe Films on
  SrTiO$_{3}$}},}\ }\href {\doibase 10.1088/0256-307X/29/3/037402} {\bibfield
  {journal} {\bibinfo  {journal} {Chinese Physics Letters}\ }\textbf {\bibinfo
  {volume} {29}},\ \bibinfo {eid} {037402} (\bibinfo {year} {2012})},\ \Eprint
  {http://arxiv.org/abs/1201.5694} {arXiv:1201.5694 [cond-mat.supr-con]}
  \BibitemShut {NoStop}%
\bibitem [{\citenamefont {Song}\ \emph {et~al.}(2011)\citenamefont {Song},
  \citenamefont {Wang}, \citenamefont {Jiang}, \citenamefont {Li},
  \citenamefont {Wang}, \citenamefont {He}, \citenamefont {Chen}, \citenamefont
  {Ma},\ and\ \citenamefont {Xue}}]{Song2011FeSe}%
  \BibitemOpen
  \bibfield  {author} {\bibinfo {author} {\bibfnamefont {Can-Li}\ \bibnamefont
  {Song}}, \bibinfo {author} {\bibfnamefont {Yi-Lin}\ \bibnamefont {Wang}},
  \bibinfo {author} {\bibfnamefont {Ye-Ping}\ \bibnamefont {Jiang}}, \bibinfo
  {author} {\bibfnamefont {Zhi}\ \bibnamefont {Li}}, \bibinfo {author}
  {\bibfnamefont {Lili}\ \bibnamefont {Wang}}, \bibinfo {author} {\bibfnamefont
  {Ke}~\bibnamefont {He}}, \bibinfo {author} {\bibfnamefont {Xi}~\bibnamefont
  {Chen}}, \bibinfo {author} {\bibfnamefont {Xu-Cun}\ \bibnamefont {Ma}}, \
  and\ \bibinfo {author} {\bibfnamefont {Qi-Kun}\ \bibnamefont {Xue}},\
  }\bibfield  {title} {\enquote {\bibinfo {title} {Molecular-beam epitaxy and
  robust superconductivity of stoichiometric fese crystalline films on bilayer
  graphene},}\ }\href {\doibase 10.1103/PhysRevB.84.020503} {\bibfield
  {journal} {\bibinfo  {journal} {Phys. Rev. B}\ }\textbf {\bibinfo {volume}
  {84}},\ \bibinfo {pages} {020503} (\bibinfo {year} {2011})}\BibitemShut
  {NoStop}%
\bibitem [{\citenamefont {Allain}\ \emph {et~al.}(2015)\citenamefont {Allain},
  \citenamefont {Kang}, \citenamefont {Banerjee},\ and\ \citenamefont
  {Kis}}]{Allain2015}%
  \BibitemOpen
  \bibfield  {author} {\bibinfo {author} {\bibfnamefont {Adrien}\ \bibnamefont
  {Allain}}, \bibinfo {author} {\bibfnamefont {Jiahao}\ \bibnamefont {Kang}},
  \bibinfo {author} {\bibfnamefont {Kaustav}\ \bibnamefont {Banerjee}}, \ and\
  \bibinfo {author} {\bibfnamefont {Andras}\ \bibnamefont {Kis}},\ }\bibfield
  {title} {\enquote {\bibinfo {title} {{Electrical contacts to two-dimensional
  semiconductors}},}\ }\href {\doibase 10.1038/nmat4452} {\bibfield  {journal}
  {\bibinfo  {journal} {Nature Materials}\ }\textbf {\bibinfo {volume} {14}},\
  \bibinfo {pages} {1195--1205} (\bibinfo {year} {2015})}\BibitemShut {NoStop}%
\bibitem [{\citenamefont {Gu}\ \emph {et~al.}(2020)\citenamefont {Gu},
  \citenamefont {Li}, \citenamefont {Wan}, \citenamefont {Li}, \citenamefont
  {Guo}, \citenamefont {Yang}, \citenamefont {Li}, \citenamefont {Zhu},
  \citenamefont {Pan}, \citenamefont {Nie} \emph {et~al.}}]{Gu2020}%
  \BibitemOpen
  \bibfield  {author} {\bibinfo {author} {\bibfnamefont {Qiangqiang}\
  \bibnamefont {Gu}}, \bibinfo {author} {\bibfnamefont {Yueying}\ \bibnamefont
  {Li}}, \bibinfo {author} {\bibfnamefont {Siyuan}\ \bibnamefont {Wan}},
  \bibinfo {author} {\bibfnamefont {Huazhou}\ \bibnamefont {Li}}, \bibinfo
  {author} {\bibfnamefont {Wei}\ \bibnamefont {Guo}}, \bibinfo {author}
  {\bibfnamefont {Huan}\ \bibnamefont {Yang}}, \bibinfo {author} {\bibfnamefont
  {Qing}\ \bibnamefont {Li}}, \bibinfo {author} {\bibfnamefont {Xiyu}\
  \bibnamefont {Zhu}}, \bibinfo {author} {\bibfnamefont {Xiaoqing}\
  \bibnamefont {Pan}}, \bibinfo {author} {\bibfnamefont {Yuefeng}\ \bibnamefont
  {Nie}},  \emph {et~al.},\ }\bibfield  {title} {\enquote {\bibinfo {title}
  {Single particle tunneling spectrum of superconducting
  nd$_{1-x}$sr$_x$nio$_2$ thin films},}\ }\href@noop {} {\bibfield  {journal}
  {\bibinfo  {journal} {Nature communications}\ }\textbf {\bibinfo {volume}
  {11}},\ \bibinfo {pages} {1--7} (\bibinfo {year} {2020})}\BibitemShut
  {NoStop}%
\bibitem [{\citenamefont {Zhang}\ \emph {et~al.}(2015)\citenamefont {Zhang},
  \citenamefont {Wang}, \citenamefont {Song}, \citenamefont {Liu},
  \citenamefont {Peng}, \citenamefont {Moler}, \citenamefont {Feng},\ and\
  \citenamefont {Wang}}]{zhang2015onset}%
  \BibitemOpen
  \bibfield  {author} {\bibinfo {author} {\bibfnamefont {Zuocheng}\
  \bibnamefont {Zhang}}, \bibinfo {author} {\bibfnamefont {Yi-Hua}\
  \bibnamefont {Wang}}, \bibinfo {author} {\bibfnamefont {Qi}~\bibnamefont
  {Song}}, \bibinfo {author} {\bibfnamefont {Chang}\ \bibnamefont {Liu}},
  \bibinfo {author} {\bibfnamefont {Rui}\ \bibnamefont {Peng}}, \bibinfo
  {author} {\bibfnamefont {KA}~\bibnamefont {Moler}}, \bibinfo {author}
  {\bibfnamefont {Donglai}\ \bibnamefont {Feng}}, \ and\ \bibinfo {author}
  {\bibfnamefont {Yayu}\ \bibnamefont {Wang}},\ }\bibfield  {title} {\enquote
  {\bibinfo {title} {Onset of the meissner effect at 65 k in fese thin film
  grown on nb-doped srtio 3 substrate},}\ }\href@noop {} {\bibfield  {journal}
  {\bibinfo  {journal} {Science bulletin}\ }\textbf {\bibinfo {volume} {60}},\
  \bibinfo {pages} {1301--1304} (\bibinfo {year} {2015})}\BibitemShut {NoStop}%
\bibitem [{\citenamefont {Xu}\ \emph {et~al.}(2019)\citenamefont {Xu},
  \citenamefont {Yu}, \citenamefont {Hui}, \citenamefont {Su}, \citenamefont
  {Cheng}, \citenamefont {Chang}, \citenamefont {Zhang}, \citenamefont {Shen},\
  and\ \citenamefont {Tian}}]{xu2019mapping}%
  \BibitemOpen
  \bibfield  {author} {\bibinfo {author} {\bibfnamefont {Ying}\ \bibnamefont
  {Xu}}, \bibinfo {author} {\bibfnamefont {Yijun}\ \bibnamefont {Yu}}, \bibinfo
  {author} {\bibfnamefont {Yuen~Yung}\ \bibnamefont {Hui}}, \bibinfo {author}
  {\bibfnamefont {Yudan}\ \bibnamefont {Su}}, \bibinfo {author} {\bibfnamefont
  {Jun}\ \bibnamefont {Cheng}}, \bibinfo {author} {\bibfnamefont {Huan-Cheng}\
  \bibnamefont {Chang}}, \bibinfo {author} {\bibfnamefont {Yuanbo}\
  \bibnamefont {Zhang}}, \bibinfo {author} {\bibfnamefont {Y~Ron}\ \bibnamefont
  {Shen}}, \ and\ \bibinfo {author} {\bibfnamefont {Chuanshan}\ \bibnamefont
  {Tian}},\ }\bibfield  {title} {\enquote {\bibinfo {title} {Mapping dynamical
  magnetic responses of ultrathin micron-size superconducting films using
  nitrogen-vacancy centers in diamond},}\ }\href {\doibase
  10.1021/acs.nanolett.9b02298} {\bibfield  {journal} {\bibinfo  {journal}
  {Nano Lett.}\ }\textbf {\bibinfo {volume} {19}},\ \bibinfo {pages}
  {5697--5702} (\bibinfo {year} {2019})}\BibitemShut {NoStop}%
\bibitem [{\citenamefont {Agarwal}\ \emph {et~al.}(2017)\citenamefont
  {Agarwal}, \citenamefont {Schmidt}, \citenamefont {Halperin}, \citenamefont
  {Oganesyan}, \citenamefont {Zar\'and}, \citenamefont {Lukin},\ and\
  \citenamefont {Demler}}]{Agarwal2017}%
  \BibitemOpen
  \bibfield  {author} {\bibinfo {author} {\bibfnamefont {Kartiek}\ \bibnamefont
  {Agarwal}}, \bibinfo {author} {\bibfnamefont {Richard}\ \bibnamefont
  {Schmidt}}, \bibinfo {author} {\bibfnamefont {Bertrand}\ \bibnamefont
  {Halperin}}, \bibinfo {author} {\bibfnamefont {Vadim}\ \bibnamefont
  {Oganesyan}}, \bibinfo {author} {\bibfnamefont {Gergely}\ \bibnamefont
  {Zar\'and}}, \bibinfo {author} {\bibfnamefont {Mikhail~D.}\ \bibnamefont
  {Lukin}}, \ and\ \bibinfo {author} {\bibfnamefont {Eugene}\ \bibnamefont
  {Demler}},\ }\bibfield  {title} {\enquote {\bibinfo {title} {Magnetic noise
  spectroscopy as a probe of local electronic correlations in two-dimensional
  systems},}\ }\href {\doibase 10.1103/PhysRevB.95.155107} {\bibfield
  {journal} {\bibinfo  {journal} {Phys. Rev. B}\ }\textbf {\bibinfo {volume}
  {95}},\ \bibinfo {pages} {155107} (\bibinfo {year} {2017})}\BibitemShut
  {NoStop}%
\bibitem [{\citenamefont {Anderson}(1958)}]{PhysRev.112.1900}%
  \BibitemOpen
  \bibfield  {author} {\bibinfo {author} {\bibfnamefont {P.~W.}\ \bibnamefont
  {Anderson}},\ }\bibfield  {title} {\enquote {\bibinfo {title} {Random-phase
  approximation in the theory of superconductivity},}\ }\href {\doibase
  10.1103/PhysRev.112.1900} {\bibfield  {journal} {\bibinfo  {journal} {Phys.
  Rev.}\ }\textbf {\bibinfo {volume} {112}},\ \bibinfo {pages} {1900--1916}
  (\bibinfo {year} {1958})}\BibitemShut {NoStop}%
\bibitem [{\citenamefont {Kulik}\ and\ \citenamefont
  {Yanson}(1974)}]{kulik1974josephson}%
  \BibitemOpen
  \bibfield  {author} {\bibinfo {author} {\bibfnamefont {I.~O.}\ \bibnamefont
  {Kulik}}\ and\ \bibinfo {author} {\bibfnamefont {I.~K.}\ \bibnamefont
  {Yanson}},\ }\bibfield  {title} {\enquote {\bibinfo {title} {The josephson
  effect in superconductive tunneling structures},}\ }\href {\doibase
  10.1119/1.1987851} {\bibfield  {journal} {\bibinfo  {journal} {American
  Journal of Physics}\ }\textbf {\bibinfo {volume} {42}},\ \bibinfo {pages}
  {799--800} (\bibinfo {year} {1974})}\BibitemShut {NoStop}%
\bibitem [{\citenamefont {Barone}\ and\ \citenamefont
  {Paterno}(1982)}]{barone1982physics}%
  \BibitemOpen
  \bibfield  {author} {\bibinfo {author} {\bibfnamefont {Antonio}\ \bibnamefont
  {Barone}}\ and\ \bibinfo {author} {\bibfnamefont {Gianfranco}\ \bibnamefont
  {Paterno}},\ }\href@noop {} {\emph {\bibinfo {title} {Physics and
  applications of the Josephson effect}}}\ (\bibinfo  {publisher} {Wiley},\
  \bibinfo {year} {1982})\BibitemShut {NoStop}%
\bibitem [{\citenamefont {Bulaevskii}\ \emph {et~al.}(1994)\citenamefont
  {Bulaevskii}, \citenamefont {Zamora}, \citenamefont {Baeriswyl},
  \citenamefont {Beck},\ and\ \citenamefont {Clem}}]{PhysRevB.50.12831}%
  \BibitemOpen
  \bibfield  {author} {\bibinfo {author} {\bibfnamefont {L.~N.}\ \bibnamefont
  {Bulaevskii}}, \bibinfo {author} {\bibfnamefont {M.}~\bibnamefont {Zamora}},
  \bibinfo {author} {\bibfnamefont {D.}~\bibnamefont {Baeriswyl}}, \bibinfo
  {author} {\bibfnamefont {H.}~\bibnamefont {Beck}}, \ and\ \bibinfo {author}
  {\bibfnamefont {John~R.}\ \bibnamefont {Clem}},\ }\bibfield  {title}
  {\enquote {\bibinfo {title} {Time-dependent equations for phase differences
  and a collective mode in josephson-coupled layered superconductors},}\ }\href
  {\doibase 10.1103/PhysRevB.50.12831} {\bibfield  {journal} {\bibinfo
  {journal} {Phys. Rev. B}\ }\textbf {\bibinfo {volume} {50}},\ \bibinfo
  {pages} {12831--12834} (\bibinfo {year} {1994})}\BibitemShut {NoStop}%
\bibitem [{\citenamefont {Sun}\ \emph {et~al.}(2020)\citenamefont {Sun},
  \citenamefont {Fogler}, \citenamefont {Basov},\ and\ \citenamefont
  {Millis}}]{Sun}%
  \BibitemOpen
  \bibfield  {author} {\bibinfo {author} {\bibfnamefont {Zhiyuan}\ \bibnamefont
  {Sun}}, \bibinfo {author} {\bibfnamefont {M.~M.}\ \bibnamefont {Fogler}},
  \bibinfo {author} {\bibfnamefont {D.~N.}\ \bibnamefont {Basov}}, \ and\
  \bibinfo {author} {\bibfnamefont {Andrew~J.}\ \bibnamefont {Millis}},\
  }\bibfield  {title} {\enquote {\bibinfo {title} {Collective modes and
  terahertz near-field response of superconductors},}\ }\href {\doibase
  10.1103/PhysRevResearch.2.023413} {\bibfield  {journal} {\bibinfo  {journal}
  {Phys. Rev. Research}\ }\textbf {\bibinfo {volume} {2}},\ \bibinfo {pages}
  {023413} (\bibinfo {year} {2020})}\BibitemShut {NoStop}%
\bibitem [{\citenamefont {Langsjoen}\ \emph {et~al.}(2012)\citenamefont
  {Langsjoen}, \citenamefont {Poudel}, \citenamefont {Vavilov},\ and\
  \citenamefont {Joynt}}]{Langsjoen}%
  \BibitemOpen
  \bibfield  {author} {\bibinfo {author} {\bibfnamefont {Luke~S.}\ \bibnamefont
  {Langsjoen}}, \bibinfo {author} {\bibfnamefont {Amrit}\ \bibnamefont
  {Poudel}}, \bibinfo {author} {\bibfnamefont {Maxim~G.}\ \bibnamefont
  {Vavilov}}, \ and\ \bibinfo {author} {\bibfnamefont {Robert}\ \bibnamefont
  {Joynt}},\ }\bibfield  {title} {\enquote {\bibinfo {title} {Qubit relaxation
  from evanescent-wave johnson noise},}\ }\href {\doibase
  10.1103/PhysRevA.86.010301} {\bibfield  {journal} {\bibinfo  {journal} {Phys.
  Rev. A}\ }\textbf {\bibinfo {volume} {86}},\ \bibinfo {pages} {010301}
  (\bibinfo {year} {2012})}\BibitemShut {NoStop}%
\bibitem [{\citenamefont {Dolgirev}\ \emph {et~al.}(2022)\citenamefont
  {Dolgirev}, \citenamefont {Chatterjee}, \citenamefont {Esterlis},
  \citenamefont {Zibrov}, \citenamefont {Lukin}, \citenamefont {Yao},\ and\
  \citenamefont {Demler}}]{Shubhayu2}%
  \BibitemOpen
  \bibfield  {author} {\bibinfo {author} {\bibfnamefont {Pavel~E.}\
  \bibnamefont {Dolgirev}}, \bibinfo {author} {\bibfnamefont {Shubhayu}\
  \bibnamefont {Chatterjee}}, \bibinfo {author} {\bibfnamefont {Ilya}\
  \bibnamefont {Esterlis}}, \bibinfo {author} {\bibfnamefont {Alexander~A.}\
  \bibnamefont {Zibrov}}, \bibinfo {author} {\bibfnamefont {Mikhail~D.}\
  \bibnamefont {Lukin}}, \bibinfo {author} {\bibfnamefont {Norman~Y.}\
  \bibnamefont {Yao}}, \ and\ \bibinfo {author} {\bibfnamefont {Eugene}\
  \bibnamefont {Demler}},\ }\bibfield  {title} {\enquote {\bibinfo {title}
  {Characterizing two-dimensional superconductivity via nanoscale noise
  magnetometry with single-spin qubits},}\ }\href {\doibase
  10.1103/PhysRevB.105.024507} {\bibfield  {journal} {\bibinfo  {journal}
  {Phys. Rev. B}\ }\textbf {\bibinfo {volume} {105}},\ \bibinfo {pages}
  {024507} (\bibinfo {year} {2022})}\BibitemShut {NoStop}%
\bibitem [{\citenamefont {Pearl}(1964)}]{pearl1964current}%
  \BibitemOpen
  \bibfield  {author} {\bibinfo {author} {\bibfnamefont {J}~\bibnamefont
  {Pearl}},\ }\bibfield  {title} {\enquote {\bibinfo {title} {Current
  distribution in superconducting films carrying quantized fluxoids},}\
  }\href@noop {} {\bibfield  {journal} {\bibinfo  {journal} {Applied Physics
  Letters}\ }\textbf {\bibinfo {volume} {5}},\ \bibinfo {pages} {65--66}
  (\bibinfo {year} {1964})}\BibitemShut {NoStop}%
\bibitem [{\citenamefont {Durst}\ and\ \citenamefont {Lee}(2000)}]{DurstLee}%
  \BibitemOpen
  \bibfield  {author} {\bibinfo {author} {\bibfnamefont {Adam~C.}\ \bibnamefont
  {Durst}}\ and\ \bibinfo {author} {\bibfnamefont {Patrick~A.}\ \bibnamefont
  {Lee}},\ }\bibfield  {title} {\enquote {\bibinfo {title} {Impurity-induced
  quasiparticle transport and universal-limit wiedemann-franz violation in
  \textit{d}-wave superconductors},}\ }\href {\doibase
  10.1103/PhysRevB.62.1270} {\bibfield  {journal} {\bibinfo  {journal} {Phys.
  Rev. B}\ }\textbf {\bibinfo {volume} {62}},\ \bibinfo {pages} {1270--1290}
  (\bibinfo {year} {2000})},\ \Eprint {http://arxiv.org/abs/cond-mat/9908182}
  {arXiv:cond-mat/9908182} \BibitemShut {NoStop}%
\bibitem [{\citenamefont {Altland}\ and\ \citenamefont
  {Simons}(2010)}]{altland2010condensed}%
  \BibitemOpen
  \bibfield  {author} {\bibinfo {author} {\bibfnamefont {Alexander}\
  \bibnamefont {Altland}}\ and\ \bibinfo {author} {\bibfnamefont {Ben~D}\
  \bibnamefont {Simons}},\ }\href@noop {} {\emph {\bibinfo {title} {Condensed
  matter field theory}}}\ (\bibinfo  {publisher} {Cambridge university press},\
  \bibinfo {year} {2010})\BibitemShut {NoStop}%
\bibitem [{\citenamefont {Khoo}\ \emph {et~al.}(2021)\citenamefont {Khoo},
  \citenamefont {Pientka},\ and\ \citenamefont {Sodemann}}]{Khoo}%
  \BibitemOpen
  \bibfield  {author} {\bibinfo {author} {\bibfnamefont {Jun~Yong}\
  \bibnamefont {Khoo}}, \bibinfo {author} {\bibfnamefont {Falko}\ \bibnamefont
  {Pientka}}, \ and\ \bibinfo {author} {\bibfnamefont {Inti}\ \bibnamefont
  {Sodemann}},\ }\bibfield  {title} {\enquote {\bibinfo {title} {The universal
  shear conductivity of fermi liquids and spinon fermi surface states and its
  detection via spin qubit noise magnetometry},}\ }\href {\doibase
  10.1088/1367-2630/ac2dab} {\bibfield  {journal} {\bibinfo  {journal} {New
  Journal of Physics}\ }\textbf {\bibinfo {volume} {23}},\ \bibinfo {pages}
  {113009} (\bibinfo {year} {2021})}\BibitemShut {NoStop}%
\bibitem [{\citenamefont {Hong}\ \emph {et~al.}(2013)\citenamefont {Hong},
  \citenamefont {Grinolds}, \citenamefont {Pham}, \citenamefont {Le~Sage},
  \citenamefont {Luan}, \citenamefont {Walsworth},\ and\ \citenamefont
  {Yacoby}}]{hong2013nanoscale}%
  \BibitemOpen
  \bibfield  {author} {\bibinfo {author} {\bibfnamefont {Sungkun}\ \bibnamefont
  {Hong}}, \bibinfo {author} {\bibfnamefont {Michael~S}\ \bibnamefont
  {Grinolds}}, \bibinfo {author} {\bibfnamefont {Linh~M}\ \bibnamefont {Pham}},
  \bibinfo {author} {\bibfnamefont {David}\ \bibnamefont {Le~Sage}}, \bibinfo
  {author} {\bibfnamefont {Lan}\ \bibnamefont {Luan}}, \bibinfo {author}
  {\bibfnamefont {Ronald~L}\ \bibnamefont {Walsworth}}, \ and\ \bibinfo
  {author} {\bibfnamefont {Amir}\ \bibnamefont {Yacoby}},\ }\bibfield  {title}
  {\enquote {\bibinfo {title} {Nanoscale magnetometry with nv centers in
  diamond},}\ }\href@noop {} {\bibfield  {journal} {\bibinfo  {journal} {MRS
  bulletin}\ }\textbf {\bibinfo {volume} {38}},\ \bibinfo {pages} {155--161}
  (\bibinfo {year} {2013})}\BibitemShut {NoStop}%
\bibitem [{\citenamefont {Grinolds}\ \emph {et~al.}(2013)\citenamefont
  {Grinolds}, \citenamefont {Hong}, \citenamefont {Maletinsky}, \citenamefont
  {Luan}, \citenamefont {Lukin}, \citenamefont {Walsworth},\ and\ \citenamefont
  {Yacoby}}]{nvsinglespin}%
  \BibitemOpen
  \bibfield  {author} {\bibinfo {author} {\bibfnamefont {M.~S.}\ \bibnamefont
  {Grinolds}}, \bibinfo {author} {\bibfnamefont {S.}~\bibnamefont {Hong}},
  \bibinfo {author} {\bibfnamefont {P.}~\bibnamefont {Maletinsky}}, \bibinfo
  {author} {\bibfnamefont {L.}~\bibnamefont {Luan}}, \bibinfo {author}
  {\bibfnamefont {M.~D.}\ \bibnamefont {Lukin}}, \bibinfo {author}
  {\bibfnamefont {R.~L.}\ \bibnamefont {Walsworth}}, \ and\ \bibinfo {author}
  {\bibfnamefont {A.}~\bibnamefont {Yacoby}},\ }\bibfield  {title} {\enquote
  {\bibinfo {title} {Nanoscale magnetic imaging of a single electron spin under
  ambient conditions},}\ }\href {http://dx.doi.org/10.1038/nphys2543}
  {\bibfield  {journal} {\bibinfo  {journal} {Nat Phys}\ }\textbf {\bibinfo
  {volume} {9}},\ \bibinfo {pages} {215--219} (\bibinfo {year}
  {2013})}\BibitemShut {NoStop}%
\bibitem [{\citenamefont {Rondin}\ \emph {et~al.}(2014)\citenamefont {Rondin},
  \citenamefont {Tetienne}, \citenamefont {Hingant}, \citenamefont {Roch},
  \citenamefont {Maletinsky},\ and\ \citenamefont {Jacques}}]{nvreview}%
  \BibitemOpen
  \bibfield  {author} {\bibinfo {author} {\bibfnamefont {L}~\bibnamefont
  {Rondin}}, \bibinfo {author} {\bibfnamefont {J-P}\ \bibnamefont {Tetienne}},
  \bibinfo {author} {\bibfnamefont {T}~\bibnamefont {Hingant}}, \bibinfo
  {author} {\bibfnamefont {J-F}\ \bibnamefont {Roch}}, \bibinfo {author}
  {\bibfnamefont {P}~\bibnamefont {Maletinsky}}, \ and\ \bibinfo {author}
  {\bibfnamefont {V}~\bibnamefont {Jacques}},\ }\bibfield  {title} {\enquote
  {\bibinfo {title} {Magnetometry with nitrogen-vacancy defects in diamond},}\
  }\href@noop {} {\bibfield  {journal} {\bibinfo  {journal} {Reports on
  Progress in Physics}\ }\textbf {\bibinfo {volume} {77}},\ \bibinfo {pages}
  {056503} (\bibinfo {year} {2014})}\BibitemShut {NoStop}%
\bibitem [{\citenamefont {{Casola}}\ \emph {et~al.}(2018)\citenamefont
  {{Casola}}, \citenamefont {{van der Sar}},\ and\ \citenamefont
  {{Yacoby}}}]{Casola}%
  \BibitemOpen
  \bibfield  {author} {\bibinfo {author} {\bibfnamefont {Francesco}\
  \bibnamefont {{Casola}}}, \bibinfo {author} {\bibfnamefont {Toeno}\
  \bibnamefont {{van der Sar}}}, \ and\ \bibinfo {author} {\bibfnamefont
  {Amir}\ \bibnamefont {{Yacoby}}},\ }\bibfield  {title} {\enquote {\bibinfo
  {title} {{Probing condensed matter physics with magnetometry based on
  nitrogen-vacancy centres in diamond}},}\ }\href {\doibase
  10.1038/natrevmats.2017.88} {\bibfield  {journal} {\bibinfo  {journal}
  {Nature Reviews Materials}\ }\textbf {\bibinfo {volume} {3}},\ \bibinfo
  {pages} {17088} (\bibinfo {year} {2018})},\ \Eprint
  {http://arxiv.org/abs/1804.08742} {arXiv:1804.08742 [cond-mat.str-el]}
  \BibitemShut {NoStop}%
\bibitem [{\citenamefont {Stano}\ \emph {et~al.}(2013)\citenamefont {Stano},
  \citenamefont {Klinovaja}, \citenamefont {Yacoby},\ and\ \citenamefont
  {Loss}}]{Stano2013}%
  \BibitemOpen
  \bibfield  {author} {\bibinfo {author} {\bibfnamefont {Peter}\ \bibnamefont
  {Stano}}, \bibinfo {author} {\bibfnamefont {Jelena}\ \bibnamefont
  {Klinovaja}}, \bibinfo {author} {\bibfnamefont {Amir}\ \bibnamefont
  {Yacoby}}, \ and\ \bibinfo {author} {\bibfnamefont {Daniel}\ \bibnamefont
  {Loss}},\ }\bibfield  {title} {\enquote {\bibinfo {title} {Local spin
  susceptibilities of low-dimensional electron systems},}\ }\href {\doibase
  10.1103/PhysRevB.88.045441} {\bibfield  {journal} {\bibinfo  {journal} {Phys.
  Rev. B}\ }\textbf {\bibinfo {volume} {88}},\ \bibinfo {pages} {045441}
  (\bibinfo {year} {2013})}\BibitemShut {NoStop}%
\bibitem [{\citenamefont {Kazmierczak}\ \emph {et~al.}(2021)\citenamefont
  {Kazmierczak}, \citenamefont {Van~Winkle}, \citenamefont {Ophus},
  \citenamefont {Bustillo}, \citenamefont {Carr}, \citenamefont {Brown},
  \citenamefont {Ciston}, \citenamefont {Taniguchi}, \citenamefont {Watanabe},\
  and\ \citenamefont {Bediako}}]{Kazmierczak_2021}%
  \BibitemOpen
  \bibfield  {author} {\bibinfo {author} {\bibfnamefont {Nathanael~P.}\
  \bibnamefont {Kazmierczak}}, \bibinfo {author} {\bibfnamefont {Madeline}\
  \bibnamefont {Van~Winkle}}, \bibinfo {author} {\bibfnamefont {Colin}\
  \bibnamefont {Ophus}}, \bibinfo {author} {\bibfnamefont {Karen~C.}\
  \bibnamefont {Bustillo}}, \bibinfo {author} {\bibfnamefont {Stephen}\
  \bibnamefont {Carr}}, \bibinfo {author} {\bibfnamefont {Hamish~G.}\
  \bibnamefont {Brown}}, \bibinfo {author} {\bibfnamefont {Jim}\ \bibnamefont
  {Ciston}}, \bibinfo {author} {\bibfnamefont {Takashi}\ \bibnamefont
  {Taniguchi}}, \bibinfo {author} {\bibfnamefont {Kenji}\ \bibnamefont
  {Watanabe}}, \ and\ \bibinfo {author} {\bibfnamefont {D.~Kwabena}\
  \bibnamefont {Bediako}},\ }\bibfield  {title} {\enquote {\bibinfo {title}
  {Strain fields in twisted bilayer graphene},}\ }\href {\doibase
  10.1038/s41563-021-00973-w} {\bibfield  {journal} {\bibinfo  {journal}
  {Nature Materials}\ }\textbf {\bibinfo {volume} {20}},\ \bibinfo {pages}
  {956–963} (\bibinfo {year} {2021})}\BibitemShut {NoStop}%
\bibitem [{\citenamefont {Jarmola}\ \emph {et~al.}(2012)\citenamefont
  {Jarmola}, \citenamefont {Acosta}, \citenamefont {Jensen}, \citenamefont
  {Chemerisov},\ and\ \citenamefont {Budker}}]{Jarmola}%
  \BibitemOpen
  \bibfield  {author} {\bibinfo {author} {\bibfnamefont {A.}~\bibnamefont
  {Jarmola}}, \bibinfo {author} {\bibfnamefont {V.~M.}\ \bibnamefont {Acosta}},
  \bibinfo {author} {\bibfnamefont {K.}~\bibnamefont {Jensen}}, \bibinfo
  {author} {\bibfnamefont {S.}~\bibnamefont {Chemerisov}}, \ and\ \bibinfo
  {author} {\bibfnamefont {D.}~\bibnamefont {Budker}},\ }\bibfield  {title}
  {\enquote {\bibinfo {title} {Temperature- and magnetic-field-dependent
  longitudinal spin relaxation in nitrogen-vacancy ensembles in diamond},}\
  }\href {\doibase 10.1103/PhysRevLett.108.197601} {\bibfield  {journal}
  {\bibinfo  {journal} {Phys. Rev. Lett.}\ }\textbf {\bibinfo {volume} {108}},\
  \bibinfo {pages} {197601} (\bibinfo {year} {2012})}\BibitemShut {NoStop}%
\bibitem [{\citenamefont {de~Guillebon}\ \emph {et~al.}(2020)\citenamefont
  {de~Guillebon}, \citenamefont {Vindolet}, \citenamefont {Roch}, \citenamefont
  {Jacques},\ and\ \citenamefont {Rondin}}]{Guillebon}%
  \BibitemOpen
  \bibfield  {author} {\bibinfo {author} {\bibfnamefont {T.}~\bibnamefont
  {de~Guillebon}}, \bibinfo {author} {\bibfnamefont {B.}~\bibnamefont
  {Vindolet}}, \bibinfo {author} {\bibfnamefont {J.-F.}\ \bibnamefont {Roch}},
  \bibinfo {author} {\bibfnamefont {V.}~\bibnamefont {Jacques}}, \ and\
  \bibinfo {author} {\bibfnamefont {L.}~\bibnamefont {Rondin}},\ }\bibfield
  {title} {\enquote {\bibinfo {title} {Temperature dependence of the
  longitudinal spin relaxation time ${T}_{1}$ of single nitrogen-vacancy
  centers in nanodiamonds},}\ }\href {\doibase 10.1103/PhysRevB.102.165427}
  {\bibfield  {journal} {\bibinfo  {journal} {Phys. Rev. B}\ }\textbf {\bibinfo
  {volume} {102}},\ \bibinfo {pages} {165427} (\bibinfo {year}
  {2020})}\BibitemShut {NoStop}%
\bibitem [{\citenamefont {Manzeli}\ \emph {et~al.}(2017)\citenamefont
  {Manzeli}, \citenamefont {Ovchinnikov}, \citenamefont {Pasquier},
  \citenamefont {Yazyev},\ and\ \citenamefont {Kis}}]{Manzeli2017}%
  \BibitemOpen
  \bibfield  {author} {\bibinfo {author} {\bibfnamefont {Sajedeh}\ \bibnamefont
  {Manzeli}}, \bibinfo {author} {\bibfnamefont {Dmitry}\ \bibnamefont
  {Ovchinnikov}}, \bibinfo {author} {\bibfnamefont {Diego}\ \bibnamefont
  {Pasquier}}, \bibinfo {author} {\bibfnamefont {Oleg~V.}\ \bibnamefont
  {Yazyev}}, \ and\ \bibinfo {author} {\bibfnamefont {Andras}\ \bibnamefont
  {Kis}},\ }\bibfield  {title} {\enquote {\bibinfo {title} {{2D transition
  metal dichalcogenides}},}\ }\href {\doibase 10.1038/natrevmats.2017.33}
  {\bibfield  {journal} {\bibinfo  {journal} {Nature Reviews Materials}\
  }\textbf {\bibinfo {volume} {2}} (\bibinfo {year} {2017}),\
  10.1038/natrevmats.2017.33}\BibitemShut {NoStop}%
\bibitem [{\citenamefont {Veyrat}\ \emph {et~al.}(2020)\citenamefont {Veyrat},
  \citenamefont {D{\'e}prez}, \citenamefont {Coissard}, \citenamefont {Li},
  \citenamefont {Gay}, \citenamefont {Watanabe}, \citenamefont {Taniguchi},
  \citenamefont {Han}, \citenamefont {Piot}, \citenamefont {Sellier},\ and\
  \citenamefont {Sac{\'e}p{\'e}}}]{Veyrat781}%
  \BibitemOpen
  \bibfield  {author} {\bibinfo {author} {\bibfnamefont {Louis}\ \bibnamefont
  {Veyrat}}, \bibinfo {author} {\bibfnamefont {Corentin}\ \bibnamefont
  {D{\'e}prez}}, \bibinfo {author} {\bibfnamefont {Alexis}\ \bibnamefont
  {Coissard}}, \bibinfo {author} {\bibfnamefont {Xiaoxi}\ \bibnamefont {Li}},
  \bibinfo {author} {\bibfnamefont {Fr{\'e}d{\'e}ric}\ \bibnamefont {Gay}},
  \bibinfo {author} {\bibfnamefont {Kenji}\ \bibnamefont {Watanabe}}, \bibinfo
  {author} {\bibfnamefont {Takashi}\ \bibnamefont {Taniguchi}}, \bibinfo
  {author} {\bibfnamefont {Zheng}\ \bibnamefont {Han}}, \bibinfo {author}
  {\bibfnamefont {Benjamin~A.}\ \bibnamefont {Piot}}, \bibinfo {author}
  {\bibfnamefont {Hermann}\ \bibnamefont {Sellier}}, \ and\ \bibinfo {author}
  {\bibfnamefont {Benjamin}\ \bibnamefont {Sac{\'e}p{\'e}}},\ }\bibfield
  {title} {\enquote {\bibinfo {title} {Helical quantum hall phase in graphene
  on srtio3},}\ }\href {\doibase 10.1126/science.aax8201} {\bibfield  {journal}
  {\bibinfo  {journal} {Science}\ }\textbf {\bibinfo {volume} {367}},\ \bibinfo
  {pages} {781--786} (\bibinfo {year} {2020})}\BibitemShut {NoStop}%
\bibitem [{\citenamefont {Chatterjee}\ \emph {et~al.}(2019)\citenamefont
  {Chatterjee}, \citenamefont {Rodriguez-Nieva},\ and\ \citenamefont
  {Demler}}]{CRD18}%
  \BibitemOpen
  \bibfield  {author} {\bibinfo {author} {\bibfnamefont {Shubhayu}\
  \bibnamefont {Chatterjee}}, \bibinfo {author} {\bibfnamefont {Joaquin~F.}\
  \bibnamefont {Rodriguez-Nieva}}, \ and\ \bibinfo {author} {\bibfnamefont
  {Eugene}\ \bibnamefont {Demler}},\ }\bibfield  {title} {\enquote {\bibinfo
  {title} {Diagnosing phases of magnetic insulators via noise magnetometry with
  spin qubits},}\ }\href {\doibase 10.1103/PhysRevB.99.104425} {\bibfield
  {journal} {\bibinfo  {journal} {Phys. Rev. B}\ }\textbf {\bibinfo {volume}
  {99}},\ \bibinfo {pages} {104425} (\bibinfo {year} {2019})}\BibitemShut
  {NoStop}%
\bibitem [{\citenamefont {{Rodriguez-Nieva}}\ \emph {et~al.}(2018)\citenamefont
  {{Rodriguez-Nieva}}, \citenamefont {{Podolsky}},\ and\ \citenamefont
  {{Demler}}}]{Joaquin_magnon_18}%
  \BibitemOpen
  \bibfield  {author} {\bibinfo {author} {\bibfnamefont {Joaquin~F.}\
  \bibnamefont {{Rodriguez-Nieva}}}, \bibinfo {author} {\bibfnamefont {Daniel}\
  \bibnamefont {{Podolsky}}}, \ and\ \bibinfo {author} {\bibfnamefont {Eugene}\
  \bibnamefont {{Demler}}},\ }\bibfield  {title} {\enquote {\bibinfo {title}
  {{Hydrodynamic sound modes and Galilean symmetry breaking in a magnon
  fluid}},}\ }\href@noop {} {\bibfield  {journal} {\bibinfo  {journal} {arXiv
  e-prints}\ ,\ \bibinfo {eid} {arXiv:1810.12333}} (\bibinfo {year} {2018})},\
  \Eprint {http://arxiv.org/abs/1810.12333} {arXiv:1810.12333
  [cond-mat.mes-hall]} \BibitemShut {NoStop}%
\bibitem [{\citenamefont {Rodriguez-Nieva}\ \emph {et~al.}(2018)\citenamefont
  {Rodriguez-Nieva}, \citenamefont {Agarwal}, \citenamefont {Giamarchi},
  \citenamefont {Halperin}, \citenamefont {Lukin},\ and\ \citenamefont
  {Demler}}]{Joaquin18}%
  \BibitemOpen
  \bibfield  {author} {\bibinfo {author} {\bibfnamefont {Joaquin~F.}\
  \bibnamefont {Rodriguez-Nieva}}, \bibinfo {author} {\bibfnamefont {Kartiek}\
  \bibnamefont {Agarwal}}, \bibinfo {author} {\bibfnamefont {Thierry}\
  \bibnamefont {Giamarchi}}, \bibinfo {author} {\bibfnamefont {Bertrand~I.}\
  \bibnamefont {Halperin}}, \bibinfo {author} {\bibfnamefont {Mikhail~D.}\
  \bibnamefont {Lukin}}, \ and\ \bibinfo {author} {\bibfnamefont {Eugene}\
  \bibnamefont {Demler}},\ }\bibfield  {title} {\enquote {\bibinfo {title}
  {Probing one-dimensional systems via noise magnetometry with single spin
  qubits},}\ }\href {\doibase 10.1103/PhysRevB.98.195433} {\bibfield  {journal}
  {\bibinfo  {journal} {Phys. Rev. B}\ }\textbf {\bibinfo {volume} {98}},\
  \bibinfo {pages} {195433} (\bibinfo {year} {2018})}\BibitemShut {NoStop}%
\bibitem [{\citenamefont {Kolkowitz}\ \emph {et~al.}(2015)\citenamefont
  {Kolkowitz}, \citenamefont {Safira}, \citenamefont {High}, \citenamefont
  {Devlin}, \citenamefont {Choi}, \citenamefont {Unterreithmeier},
  \citenamefont {Patterson}, \citenamefont {Zibrov}, \citenamefont
  {Manucharyan}, \citenamefont {Park} \emph {et~al.}}]{Kolkowitz}%
  \BibitemOpen
  \bibfield  {author} {\bibinfo {author} {\bibfnamefont {S}~\bibnamefont
  {Kolkowitz}}, \bibinfo {author} {\bibfnamefont {A}~\bibnamefont {Safira}},
  \bibinfo {author} {\bibfnamefont {AA}~\bibnamefont {High}}, \bibinfo {author}
  {\bibfnamefont {RC}~\bibnamefont {Devlin}}, \bibinfo {author} {\bibfnamefont
  {S}~\bibnamefont {Choi}}, \bibinfo {author} {\bibfnamefont {QP}~\bibnamefont
  {Unterreithmeier}}, \bibinfo {author} {\bibfnamefont {D}~\bibnamefont
  {Patterson}}, \bibinfo {author} {\bibfnamefont {AS}~\bibnamefont {Zibrov}},
  \bibinfo {author} {\bibfnamefont {VE}~\bibnamefont {Manucharyan}}, \bibinfo
  {author} {\bibfnamefont {H}~\bibnamefont {Park}},  \emph {et~al.},\
  }\bibfield  {title} {\enquote {\bibinfo {title} {Probing johnson noise and
  ballistic transport in normal metals with a single-spin qubit},}\ }\href@noop
  {} {\bibfield  {journal} {\bibinfo  {journal} {Science}\ }\textbf {\bibinfo
  {volume} {347}},\ \bibinfo {pages} {1129--1132} (\bibinfo {year}
  {2015})}\BibitemShut {NoStop}%
\bibitem [{\citenamefont {{Hsieh}}\ \emph {et~al.}(2019)\citenamefont
  {{Hsieh}}, \citenamefont {{Bhattacharyya}}, \citenamefont {{Zu}},
  \citenamefont {{Mittiga}}, \citenamefont {{Smart}}, \citenamefont
  {{Machado}}, \citenamefont {{Kobrin}}, \citenamefont {{H{\"o}hn}},
  \citenamefont {{Rui}}, \citenamefont {{Kamrani}}, \citenamefont
  {{Chatterjee}}, \citenamefont {{Choi}}, \citenamefont {{Zaletel}},
  \citenamefont {{Struzhkin}}, \citenamefont {{Moore}}, \citenamefont
  {{Levitas}}, \citenamefont {{Jeanloz}},\ and\ \citenamefont
  {{Yao}}}]{Satcher}%
  \BibitemOpen
  \bibfield  {author} {\bibinfo {author} {\bibfnamefont {S.}~\bibnamefont
  {{Hsieh}}}, \bibinfo {author} {\bibfnamefont {P.}~\bibnamefont
  {{Bhattacharyya}}}, \bibinfo {author} {\bibfnamefont {C.}~\bibnamefont
  {{Zu}}}, \bibinfo {author} {\bibfnamefont {T.}~\bibnamefont {{Mittiga}}},
  \bibinfo {author} {\bibfnamefont {T.~J.}\ \bibnamefont {{Smart}}}, \bibinfo
  {author} {\bibfnamefont {F.}~\bibnamefont {{Machado}}}, \bibinfo {author}
  {\bibfnamefont {B.}~\bibnamefont {{Kobrin}}}, \bibinfo {author}
  {\bibfnamefont {T.~O.}\ \bibnamefont {{H{\"o}hn}}}, \bibinfo {author}
  {\bibfnamefont {N.~Z.}\ \bibnamefont {{Rui}}}, \bibinfo {author}
  {\bibfnamefont {M.}~\bibnamefont {{Kamrani}}}, \bibinfo {author}
  {\bibfnamefont {S.}~\bibnamefont {{Chatterjee}}}, \bibinfo {author}
  {\bibfnamefont {S.}~\bibnamefont {{Choi}}}, \bibinfo {author} {\bibfnamefont
  {M.}~\bibnamefont {{Zaletel}}}, \bibinfo {author} {\bibfnamefont {V.~V.}\
  \bibnamefont {{Struzhkin}}}, \bibinfo {author} {\bibfnamefont {J.~E.}\
  \bibnamefont {{Moore}}}, \bibinfo {author} {\bibfnamefont {V.~I.}\
  \bibnamefont {{Levitas}}}, \bibinfo {author} {\bibfnamefont {R.}~\bibnamefont
  {{Jeanloz}}}, \ and\ \bibinfo {author} {\bibfnamefont {N.~Y.}\ \bibnamefont
  {{Yao}}},\ }\bibfield  {title} {\enquote {\bibinfo {title} {{Imaging stress
  and magnetism at high pressures using a nanoscale quantum sensor}},}\ }\href
  {\doibase 10.1126/science.aaw4352} {\bibfield  {journal} {\bibinfo  {journal}
  {Science}\ }\textbf {\bibinfo {volume} {366}},\ \bibinfo {pages} {1349--1354}
  (\bibinfo {year} {2019})},\ \Eprint {http://arxiv.org/abs/1812.08796}
  {arXiv:1812.08796 [cond-mat.mes-hall]} \BibitemShut {NoStop}%
\bibitem [{\citenamefont {Rustagi}\ \emph {et~al.}(2020)\citenamefont
  {Rustagi}, \citenamefont {Bertelli}, \citenamefont {van~der Sar},\ and\
  \citenamefont {Upadhyaya}}]{Rustagi}%
  \BibitemOpen
  \bibfield  {author} {\bibinfo {author} {\bibfnamefont {Avinash}\ \bibnamefont
  {Rustagi}}, \bibinfo {author} {\bibfnamefont {Iacopo}\ \bibnamefont
  {Bertelli}}, \bibinfo {author} {\bibfnamefont {Toeno}\ \bibnamefont {van~der
  Sar}}, \ and\ \bibinfo {author} {\bibfnamefont {Pramey}\ \bibnamefont
  {Upadhyaya}},\ }\bibfield  {title} {\enquote {\bibinfo {title} {Sensing
  chiral magnetic noise via quantum impurity relaxometry},}\ }\href {\doibase
  10.1103/PhysRevB.102.220403} {\bibfield  {journal} {\bibinfo  {journal}
  {Phys. Rev. B}\ }\textbf {\bibinfo {volume} {102}},\ \bibinfo {pages}
  {220403} (\bibinfo {year} {2020})}\BibitemShut {NoStop}%
\bibitem [{\citenamefont {{Wang}}\ \emph {et~al.}(2020)\citenamefont {{Wang}},
  \citenamefont {{Zhang}}, \citenamefont {{McLaughlin}}, \citenamefont
  {{Flebus}}, \citenamefont {{Huang}}, \citenamefont {{Xiao}}, \citenamefont
  {{Fullerton}}, \citenamefont {{Tserkovnyak}},\ and\ \citenamefont
  {{Du}}}]{Du}%
  \BibitemOpen
  \bibfield  {author} {\bibinfo {author} {\bibfnamefont {Hailong}\ \bibnamefont
  {{Wang}}}, \bibinfo {author} {\bibfnamefont {Shu}\ \bibnamefont {{Zhang}}},
  \bibinfo {author} {\bibfnamefont {Nathan~J.}\ \bibnamefont {{McLaughlin}}},
  \bibinfo {author} {\bibfnamefont {Benedetta}\ \bibnamefont {{Flebus}}},
  \bibinfo {author} {\bibfnamefont {Mengqi}\ \bibnamefont {{Huang}}}, \bibinfo
  {author} {\bibfnamefont {Yuxuan}\ \bibnamefont {{Xiao}}}, \bibinfo {author}
  {\bibfnamefont {Eric~E.}\ \bibnamefont {{Fullerton}}}, \bibinfo {author}
  {\bibfnamefont {Yaroslav}\ \bibnamefont {{Tserkovnyak}}}, \ and\ \bibinfo
  {author} {\bibfnamefont {Chunhui~Rita}\ \bibnamefont {{Du}}},\ }\bibfield
  {title} {\enquote {\bibinfo {title} {{Quantum Sensing of Spin Transport
  Properties of an Antiferromagnetic Insulator}},}\ }\href@noop {} {\bibfield
  {journal} {\bibinfo  {journal} {arXiv e-prints}\ ,\ \bibinfo {eid}
  {arXiv:2011.03905}} (\bibinfo {year} {2020})},\ \Eprint
  {http://arxiv.org/abs/2011.03905} {arXiv:2011.03905 [cond-mat.mes-hall]}
  \BibitemShut {NoStop}%
\bibitem [{\citenamefont {Andersen}\ \emph {et~al.}(2019)\citenamefont
  {Andersen}, \citenamefont {Dwyer}, \citenamefont {Sanchez-Yamagishi},
  \citenamefont {Rodriguez-Nieva}, \citenamefont {Agarwal}, \citenamefont
  {Watanabe}, \citenamefont {Taniguchi}, \citenamefont {Demler}, \citenamefont
  {Kim}, \citenamefont {Park} \emph {et~al.}}]{AndersenDwyer}%
  \BibitemOpen
  \bibfield  {author} {\bibinfo {author} {\bibfnamefont {Trond~I}\ \bibnamefont
  {Andersen}}, \bibinfo {author} {\bibfnamefont {Bo~L}\ \bibnamefont {Dwyer}},
  \bibinfo {author} {\bibfnamefont {Javier~D}\ \bibnamefont
  {Sanchez-Yamagishi}}, \bibinfo {author} {\bibfnamefont {Joaquin~F}\
  \bibnamefont {Rodriguez-Nieva}}, \bibinfo {author} {\bibfnamefont {Kartiek}\
  \bibnamefont {Agarwal}}, \bibinfo {author} {\bibfnamefont {Kenji}\
  \bibnamefont {Watanabe}}, \bibinfo {author} {\bibfnamefont {Takashi}\
  \bibnamefont {Taniguchi}}, \bibinfo {author} {\bibfnamefont {Eugene~A}\
  \bibnamefont {Demler}}, \bibinfo {author} {\bibfnamefont {Philip}\
  \bibnamefont {Kim}}, \bibinfo {author} {\bibfnamefont {Hongkun}\ \bibnamefont
  {Park}},  \emph {et~al.},\ }\bibfield  {title} {\enquote {\bibinfo {title}
  {Electron-phonon instability in graphene revealed by global and local noise
  probes},}\ }\href@noop {} {\bibfield  {journal} {\bibinfo  {journal}
  {Science}\ }\textbf {\bibinfo {volume} {364}},\ \bibinfo {pages} {154--157}
  (\bibinfo {year} {2019})}\BibitemShut {NoStop}%
\bibitem [{\citenamefont {Myers}\ \emph {et~al.}(2017)\citenamefont {Myers},
  \citenamefont {Ariyaratne},\ and\ \citenamefont
  {Jayich}}]{PhysRevLett.118.197201}%
  \BibitemOpen
  \bibfield  {author} {\bibinfo {author} {\bibfnamefont {B.~A.}\ \bibnamefont
  {Myers}}, \bibinfo {author} {\bibfnamefont {A.}~\bibnamefont {Ariyaratne}}, \
  and\ \bibinfo {author} {\bibfnamefont {A.~C.~Bleszynski}\ \bibnamefont
  {Jayich}},\ }\bibfield  {title} {\enquote {\bibinfo {title} {Double-quantum
  spin-relaxation limits to coherence of near-surface nitrogen-vacancy
  centers},}\ }\href {\doibase 10.1103/PhysRevLett.118.197201} {\bibfield
  {journal} {\bibinfo  {journal} {Phys. Rev. Lett.}\ }\textbf {\bibinfo
  {volume} {118}},\ \bibinfo {pages} {197201} (\bibinfo {year}
  {2017})}\BibitemShut {NoStop}%
\bibitem [{\citenamefont {{Sahay}}\ \emph {et~al.}(2021)\citenamefont
  {{Sahay}}, \citenamefont {{Hsieh}}, \citenamefont {{Parsonnet}},
  \citenamefont {{Martin}}, \citenamefont {{Ramesh}}, \citenamefont {{Yao}},\
  and\ \citenamefont {{Chatterjee}}}]{Sahay2021}%
  \BibitemOpen
  \bibfield  {author} {\bibinfo {author} {\bibfnamefont {Rahul}\ \bibnamefont
  {{Sahay}}}, \bibinfo {author} {\bibfnamefont {Satcher}\ \bibnamefont
  {{Hsieh}}}, \bibinfo {author} {\bibfnamefont {Eric}\ \bibnamefont
  {{Parsonnet}}}, \bibinfo {author} {\bibfnamefont {Lane~W.}\ \bibnamefont
  {{Martin}}}, \bibinfo {author} {\bibfnamefont {Ramamoorthy}\ \bibnamefont
  {{Ramesh}}}, \bibinfo {author} {\bibfnamefont {Norman~Y.}\ \bibnamefont
  {{Yao}}}, \ and\ \bibinfo {author} {\bibfnamefont {Shubhayu}\ \bibnamefont
  {{Chatterjee}}},\ }\bibfield  {title} {\enquote {\bibinfo {title} {{Noise
  Electrometry of Polar and Dielectric Materials}},}\ }\href@noop {} {\bibfield
   {journal} {\bibinfo  {journal} {arXiv e-prints}\ ,\ \bibinfo {eid}
  {arXiv:2111.09315}} (\bibinfo {year} {2021})},\ \Eprint
  {http://arxiv.org/abs/2111.09315} {arXiv:2111.09315 [cond-mat.mtrl-sci]}
  \BibitemShut {NoStop}%
\bibitem [{\citenamefont {Gottscholl}\ \emph {et~al.}(2020)\citenamefont
  {Gottscholl}, \citenamefont {Kianinia}, \citenamefont {Soltamov},
  \citenamefont {Orlinskii}, \citenamefont {Mamin}, \citenamefont {Bradac},
  \citenamefont {Kasper}, \citenamefont {Krambrock}, \citenamefont {Sperlich},
  \citenamefont {Toth},\ and\ \citenamefont {et~al.}}]{gottscholl2019room}%
  \BibitemOpen
  \bibfield  {author} {\bibinfo {author} {\bibfnamefont {Andreas}\ \bibnamefont
  {Gottscholl}}, \bibinfo {author} {\bibfnamefont {Mehran}\ \bibnamefont
  {Kianinia}}, \bibinfo {author} {\bibfnamefont {Victor}\ \bibnamefont
  {Soltamov}}, \bibinfo {author} {\bibfnamefont {Sergei}\ \bibnamefont
  {Orlinskii}}, \bibinfo {author} {\bibfnamefont {Georgy}\ \bibnamefont
  {Mamin}}, \bibinfo {author} {\bibfnamefont {Carlo}\ \bibnamefont {Bradac}},
  \bibinfo {author} {\bibfnamefont {Christian}\ \bibnamefont {Kasper}},
  \bibinfo {author} {\bibfnamefont {Klaus}\ \bibnamefont {Krambrock}}, \bibinfo
  {author} {\bibfnamefont {Andreas}\ \bibnamefont {Sperlich}}, \bibinfo
  {author} {\bibfnamefont {Milos}\ \bibnamefont {Toth}}, \ and\ \bibinfo
  {author} {\bibnamefont {et~al.}},\ }\bibfield  {title} {\enquote {\bibinfo
  {title} {Initialization and read-out of intrinsic spin defects in a van der
  waals crystal at room temperature},}\ }\href {\doibase
  10.1038/s41563-020-0619-6} {\bibfield  {journal} {\bibinfo  {journal} {Nature
  Materials}\ }\textbf {\bibinfo {volume} {19}},\ \bibinfo {pages} {540–545}
  (\bibinfo {year} {2020})}\BibitemShut {NoStop}%
\bibitem [{\citenamefont {{Tetienne}}(2021)}]{Tetienne}%
  \BibitemOpen
  \bibfield  {author} {\bibinfo {author} {\bibfnamefont {J.~P.}\ \bibnamefont
  {{Tetienne}}},\ }\bibfield  {title} {\enquote {\bibinfo {title} {{Quantum
  sensors go flat}},}\ }\href {\doibase 10.1038/s41567-021-01338-5} {\bibfield
  {journal} {\bibinfo  {journal} {Nature Physics}\ }\textbf {\bibinfo {volume}
  {17}},\ \bibinfo {pages} {1074--1075} (\bibinfo {year} {2021})}\BibitemShut
  {NoStop}%
\end{thebibliography}%
\end{document}